\documentclass[aps,prb,preprint,superscriptaddress]{revtex4-1}
\usepackage{amsmath}
\usepackage{textcomp}
\usepackage{units}
\usepackage{graphicx, color, psfrag}
\usepackage{float}

\begin{document}


\title{Coherent Acoustic Perturbation of Second-Harmonic-Generation in NiO}



\author{L. Huber}
\email[]{huberluc@phys.ethz.ch}
\affiliation{Institute for Quantum Electronics, ETH Z\"urich, 8093 Zurich, Switzerland}
\author{A. Ferrer}
\affiliation{Institute for Quantum Electronics, ETH Z\"urich, 8093 Zurich, Switzerland}
\affiliation{Swiss Light Source, Paul Scherrer Institut, 5232 Villigen PSI, Switzerland}
\author{T. Kubacka}
\affiliation{Institute for Quantum Electronics, ETH Z\"urich, 8093 Zurich, Switzerland}
\author{T. Huber}
\affiliation{Institute for Quantum Electronics, ETH Z\"urich, 8093 Zurich, Switzerland}
\author{C. Dornes}
\affiliation{Institute for Quantum Electronics, ETH Z\"urich, 8093 Zurich, Switzerland}
\author{T. Sato}
\affiliation{RIKEN SPring-8 Center, Harima Institute, Sayo, Hyogo 679-5148, Japan}
\author{K. Ogawa}
\affiliation{RIKEN SPring-8 Center, Harima Institute, Sayo, Hyogo 679-5148, Japan}
\author{K. Tono}
\affiliation{RIKEN SPring-8 Center, Harima Institute, Sayo, Hyogo 679-5148, Japan}
\author{T. Katayama}
\affiliation{RIKEN SPring-8 Center, Harima Institute, Sayo, Hyogo 679-5148, Japan}
\author{Y. Inubushi}
\affiliation{RIKEN SPring-8 Center, Harima Institute, Sayo, Hyogo 679-5148, Japan}
\author{M. Yabashi}
\affiliation{RIKEN SPring-8 Center, Harima Institute, Sayo, Hyogo 679-5148, Japan}
\author{Yoshikazu Tanaka}
\affiliation{RIKEN SPring-8 Center, Harima Institute, Sayo, Hyogo 679-5148, Japan}
\author{P. Beaud}
\affiliation{Swiss Light Source, Paul Scherrer Institut, 5232 Villigen PSI, Switzerland}
\author{M. Fiebig}
\affiliation{Department of Materials, ETH Z\"urich, 8093 Zurich, Switzerland}
\author{V. Scagnoli}
\affiliation{Institute for Quantum Electronics, ETH Z\"urich, 8093 Zurich, Switzerland}
\affiliation{Laboratory for Mesoscopic Systems, Department of Materials, ETH Z\"urich, 8093 Zurich, Switzerland} 
\affiliation{Laboratory for Micro- and Nanotechnology, Paul Scherrer Institute, 5232 Villigen PSI, Switzerland}
\author{U. Staub}
\affiliation{Swiss Light Source, Paul Scherrer Institut, 5232 Villigen PSI, Switzerland}
\author{S. L. Johnson}
\affiliation{Institute for Quantum Electronics, ETH Z\"urich, 8093 Zurich, Switzerland}


\date{\today}

\begin{abstract} 
We investigate the structural and magnetic origins of the unusual ultrafast second-harmonic-generation (SHG) response of femtosecond-laser-excited nickel oxide (NiO) previously attributed to oscillatory reorientation dynamics of the magnetic structure induced by d-d excitations. Using time-resolved x-ray diffraction from the ($\frac{3}{2}\,\frac{3}{2}\,\frac{3}{2}$) magnetic planes, we show that changes in the magnitude of the magnetic structure factor following ultrafast optical excitation are limited to $\Delta \left< F_m\right>/\left< F_m\right> = 1.5\%$ in the first 30 ps. 
An extended investigation of the ultrafast SHG response reveals a strong dependence on wavelength as well as characteristic echoes, both of which give evidence for an acoustic origin of the dynamics.
We therefore propose an alternative mechanism for the SHG response based on perturbations of the nonlinear susceptibility via optically induced strain in a spatially confined medium. In this model, the two observed oscillation periods can be understood as the times required for an acoustic strain wave to traverse one coherence length of the SHG process in either the collinear or anti-collinear geometries.
\end{abstract}


\maketitle


As materials for spintronic applications, antiferromagnets (AFMs) provide the advantageous property of potentially faster spin switching times compared to ferromagnets \cite{fmresonance, afmresonance, RasingTmFeO3}. By exploiting the exchange bias interaction that arises on an interface to a ferromagnetic material \cite{exchangebias}, AFMs could act as fast switches in future data storage devices. Due to its high N\'{e}el temperature $T_N = 523$ K, the AFM NiO is a promising candidate for such an application provided that the spin order can indeed be quickly and efficiently switched.  

Controlling the antiferromagnetic order in NiO appeared to become feasible after it was reported that ultrafast optical excitation could lead to a change in the magnetic anisotropy potential, inducing 90$^{\circ}$ flips in a substantial fraction of the spin population and a rotation of the macroscopic order parameter alongside a sudden decrease in spin order \cite{SHGpumpprobe, SHGpumpprobe2,SHGpulselength}. Time-resolved SHG studies on bulk NiO further suggested that the dynamics were highly sensitive to the intensity and duration of the excitation pulse, resulting in transient redirection of spins oscillating in either of two distinct directions with frequencies of 1 and 55 GHz, respectively \cite{SHGpulselength}. 
This model was supported by the agreement of the observed frequencies with the frequencies associated with differences in anisotropy energy to the easy axis ground state known from neutron diffraction \cite{neutrondiff}. Similar mechanisms leading to spin reorientation after optical excitation have been described for AFM compounds exhibiting a net magnetization \cite{RasingTmFeO3}.  These materials, however, show a strongly temperature-dependent magnetic anisotropy that is not seen in NiO. The observations of AFM dynamics in NiO have so far only been made using resonantly enhanced SHG of $EH$-type, involving a magnetic-dipole transition. This is an elegant but  indirect measure of magnetism in NiO which poses significant experimental challenges \cite{SHGinNiO}.

Similar optical excitation levels in absorbing solids also generate coherent strain waves with frequency components similar to some of those observed in the SHG measurements \cite{StrainWave}.  For NiO in particular, Bosco et al.\ observed oscillations in time-resolved linear reflectivity measurements and Takahara et al.\ recently remarked at the close resemblance between frequencies of these coherent oscillations and the higher frequency oscillations reported for the SHG response \cite{BoscoNiO, StrainNiO}. It remains, however, unclear how such coherent strain waves can adequately explain all features observed in SHG measurements.  In particular the magnitude of the changes in SHG are significantly larger than those seen in reflectivity, and the dependence on fluence appears to be substantially different. Linear reflectivity measurements also show no evidence of the 1 GHz oscillations seen in SHG.  

By performing additional measurements and simulations, here we demonstrate that the dynamics observed in the SHG response to ultrafast optical excitation can be explained self-consistently as the result of coherent strain propagation.
For this we used time-resolved x-ray diffraction to directly measure the evolution of spin order after excitation in this system and performed extended SHG studies that validate and quantify the strain wave induced origin of the SHG dynamics in NiO. 

\section{Structural, magnetic and domain properties in NiO}

Above its N\'{e}el temperature, NiO is paramagnetic and is found in a rock-salt crystal structure. Below $T_N$, $\rm{Ni}^{2+}$ spins align ferromagnetically within $(111)$ planes, pointing along one of three equivalent $\left\{ 11\overline{2} \right\}$ directions.  The magnetization direction of neighboring $(111)$ planes is antiparallel.
Magnetostriction leads to small structural contractions along $(111)$, giving rise to four possible structural domain orientations derived from the cubic paramagnetic structure \cite{roth, SHGdomains}. The current model of anisotropy change following ultrafast optical excitation in this compound assumes a partial reorientation of the $\left\{ 11\overline{2} \right\}$ spin population, either along $(111)$ or $\left\{1\overline{1}0 \right\}$, depending on specific excitation  parameters \cite{SHGpumpprobe, SHGpumpprobe2, SHGpulselength}.

In order to disentangle influences arising from the complex domain structure, we prepared three different NiO specimens with $(111)$ surface orientation that underwent different annealing and polishing processes. We prepared samples of 36, 45 and 50 \textmu m thickness. The latter two showed bright green color and structural domain sizes of 10-100 \textmu m as well as single spin domains, which was verified using birefringence microscopy and SHG measurements \cite{SHGdomains}. The 36 \textmu m sample showed evenly distributed structural domains smaller than 10 \textmu m and was of brownish green color hinting at a deviation from ideal stoichiometry \cite{NiOoptprop}. 

\section{Time-resolved x-ray diffraction}

In order to quantify the response of the magnetic order to optical excitation, we used time-resolved non-resonant magnetic x-ray diffraction to measure the sublattice magnetization as a function of time after absorption of a femtosecond optical pulse. AFMs like NiO are ideal systems to be studied with magnetic diffraction, as the alternating magnetic moments form a sublattice with half the periodicity of the structural lattice, giving rise to additional solitary magnetic reflections in reciprocal space\cite{roth}. 

The intensity of these magnetic diffraction peaks is proportional to the square of the magnetic structure factor $\left< \textbf{F}_m(\rm{\textbf{Q}})\right>$ where $\textbf{Q}$ is the momentum transfer of the diffraction peak \cite{bergevinNiO}. 
\begin{equation}
\left< {F}_m(\textbf{Q})\right> =  - r_{0} \frac{i h \nu}{m_{e} c^2}\left<{M}_m(\rm{\textbf{Q}})\right>.
\label{eq:structurefactor}
\end{equation}
Here, $m_e$ and $r_0$ are the  mass and classical radius of the electron, $c$ is the speed of light and $h \nu$ gives the photon energy. The magnetic scattering amplitude $ {M}_m(\rm{\textbf{Q}})$ is given by
\begin{equation}
\left< M(\mathbf{Q}) \right> = \frac{1}{2} \mathbf{L}(\mathbf{Q}) \cdot \mathbf{A} + \mathbf{S}(\mathbf{Q}) \cdot \mathbf{B} \label{eq:M}
\end{equation}
where $\mathbf{A}$ and $\mathbf{B}$ are geometric factors that depend on the scattering geometry and on polarization for both the incident and scattered x-rays, $\textbf{S}(\textbf{Q})$ is the Fourier component at $\mathbf{Q}$ of the spin density,  and $\textbf{L}(\textbf{Q})$ is the Fourier component at $\mathbf{Q}$ of a function related to orbital contributions to the magnetization \cite{blumegibbs}.   

In 1972 Brunel and Bergevin used NiO to demonstrate for the first time the feasibility of magnetic x-ray diffraction \cite{bergevin1972}. Since then, the brightness of light sources increased tremendously, but the low efficiency of magnetic scattering still impedes time-resolved measurements with ps resolution at conventional light sources. For this reason the experiment was carried out at beamline 3 of the x-ray free electron laser (XFEL) SACLA, Japan \cite{sacla}. 

\begin{figure}[htbp]
	\centering
\includegraphics[width=.48\textwidth]{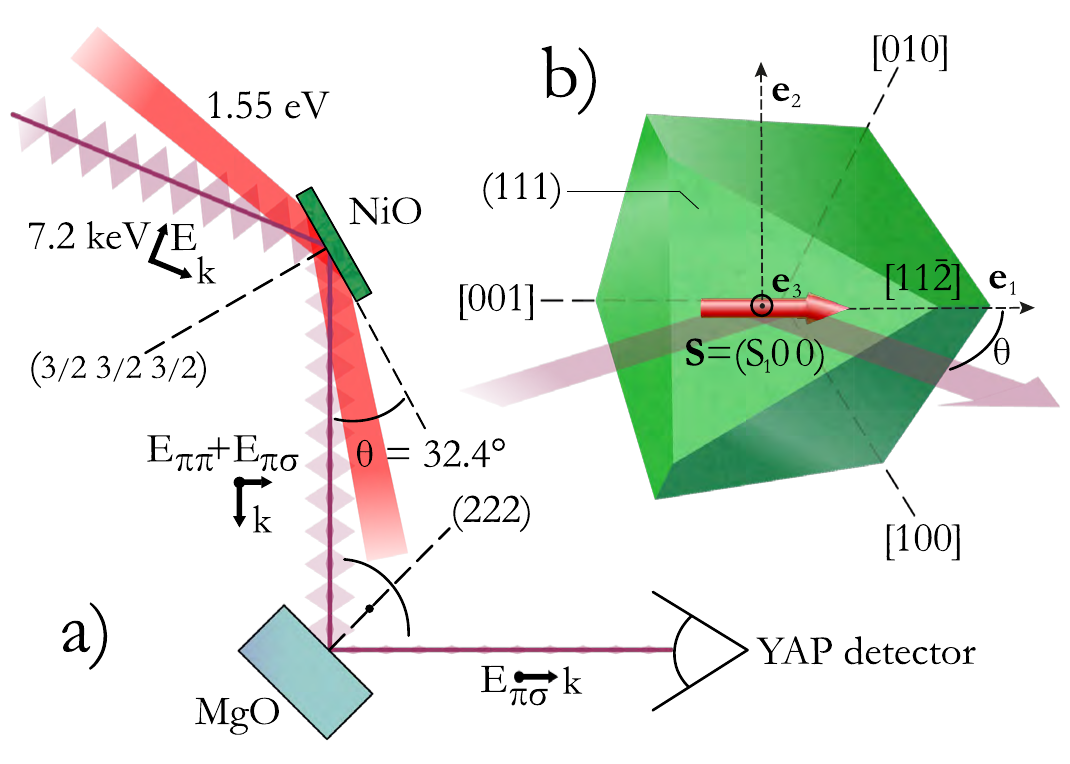}
\caption{Panel (a) depicts the experimental diffraction setup of the horizontal scattering geometry at beamline 3 of the XFEL SACLA. An MgO crystal at a Bragg angle of $45^{\circ}$ was used to select outgoing $\sigma$ radiation that was detected using an yttrium aluminium perovskite detector (YAP). Panel (b) shows the spin orientation in the antiferromagnetic phase of NiO for a single spin domain of a $(111)$-cut crystal where the spin points along $\left\{ 11\overline{2} \right\}$.}
	\label{fig:NiO_xrayAB}
\end{figure}

\begin{figure}[htbp]
	\centering
	\includegraphics[width=.480\textwidth]{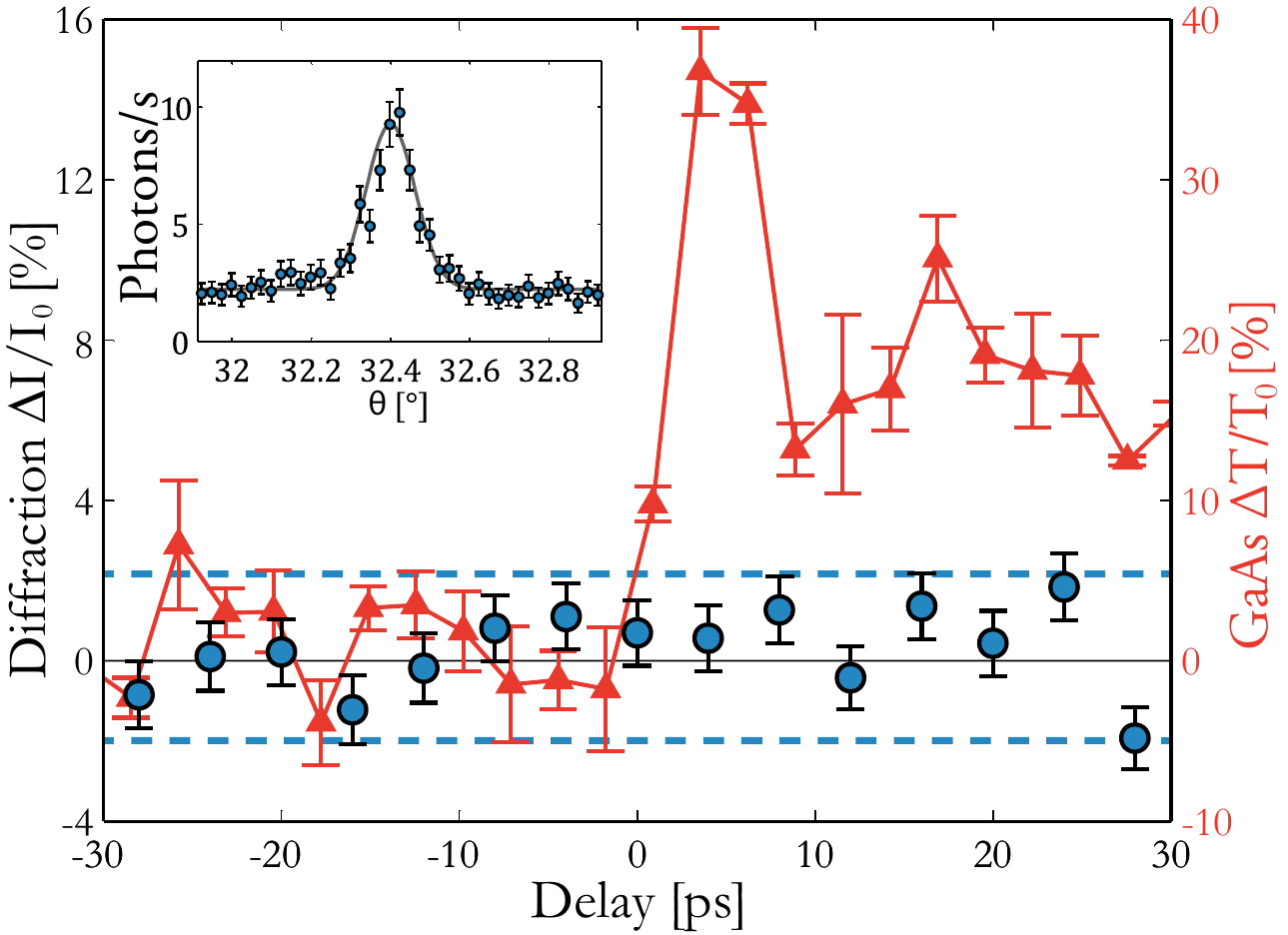}
\caption{The time resolved diffraction measurements on NiO are shown as blue dots. Broken blue lines indicate two times the mean standard error of photon counting statistics, approximately $2\%$. We did not observe a relative change in diffraction intensity exceeding this value. Temporal and spatial overlap were verified using an x-ray pump / optical transmission probe experiment on GaAs depicted as orange triangles. The inset shows a scan of the NiO Bragg angle $\theta$ in the vicinity of the ($\frac{3}{2}$ $\frac{3}{2}$ $\frac{3}{2}$) reflection. The time scan was carried out at its peak position of $\theta = 32.4^{\circ}$. }
	\label{fig:NiO_xrayC}
\end{figure}

The 45 \textmu m thick (111)-oriented NiO crystal was mounted on a multi-axis diffractometer in horizontal scattering geometry as depicted in Fig.\ \ref{fig:NiO_xrayAB}(a)  ($\pi$ incident x-ray geometry).
To measure the intensity of magnetic diffraction from the ($\frac{3}{2}$ $\frac{3}{2}$ $\frac{3}{2}$) planes, x-ray pulses from the XFEL were set to an average photon energy of 7.2 keV, with a full-width-at-half-maximum (FWHM) pulse duration below 50 fs \cite{saclapulselength, saclatiming} and an estimated spectral bandwidth of 50 eV. At a repetition rate of 10 Hz the total pulse energy was approximately 180 \textmu J. The Bragg angle for the ($\frac{3}{2}$ $\frac{3}{2}$ $\frac{3}{2}$)  planes at this photon energy is $32.4^\circ$.  The x-ray footprint on the crystal was (0.40$\times$0.75) mm$^2$.
Polarization analysis using a magnesium oxide analyzer crystal in $\pi\sigma$-configuration efficiently suppressed the charge-scattered background, allowing better measurement of the the weak magnetic signal.
In this configuration the ($\frac{3}{2}$ $\frac{3}{2}$ $\frac{3}{2}$) reflection gives the strongest signal relative to other magnetic reflections, yielding 0.9 photons per XFEL-pulse, as shown in the inset of Fig.\ \ref{fig:NiO_xrayC}.

To test for possible contributions from the XFEL second harmonic diffracting from the $(333)$ structural planes, we inserted a Si filter with a nominal thickness of 100 \textmu m  into the beam path.
This thickness of Si transmits 14\% of  x-ray radiation at 7.2 keV, but for any potential second harmonic contributions at 14.4 keV the transmission is 77\%.  We observed a transmission of the diffracted signal of $(12.2 \pm 0.6)$\%, indicating that second harmonic contributions from the (333) structurally allowed reflection are not significant in this experiment.  

To excite the sample, an amplified Ti-sapphire laser system synchronized to the XFEL provided pump pulses centered at 800 nm wavelength with 50 fs FWHM duration. The timing jitter of less than 1 ps exceeded our requirements in order to resolve a potential 20 ps oscillation period, corresponding to a 50 GHz response\cite{saclatiming}.  The pump pulses were focused onto a (0.7$\times$1.7) mm$^2$ (FWHM) spot on the sample at an incidence angle of $24.4^\circ$, which is close to Brewster's angle and $8^\circ$ degrees from the x-ray beam.  The incident excitation fluence was 37 mJ/cm$^2$, leading to an excitation density of $0.5 \cdot 10^{20}$ cm$^{-3}$ near the sample surface.  
As can be seen in section 4, these pump conditions lead to large magnitude SHG dynamics, which were interpreted as a reorientation of spins in the $\left[111\right]$ direction \cite{SHGpumpprobe}. Spatial and temporal overlap of the pump and probe pulses were verified by measuring the x-ray induced optical transmission changes of a GaAs wafer temporarily inserted into the sample position\cite{GaAs}.  The intensity attenuation depth was 24 \textmu m and 26 \textmu m for the x-rays and the pump beam, respectively.
 We measured the intensity of the ($\frac{3}{2}$ $\frac{3}{2}$ $\frac{3}{2}$) peak at room temperature as a function of relative pump-probe delay time over a range of $-30$ to $30$ ps.
As shown in Fig.\ 2, the measured relative changes in the diffraction intensity did not exceed a value of two times the mean error of photon counting statistics of 2\% over the first 30 ps after excitation. 

\section{Discussion of the diffraction results}

The absence of a change in magnetic diffraction intensity exceeding $2\%$ stands in contrast to the SHG response where a drop of more than $20\%$ is seen with similar excitation conditions (see Fig.\ \ref{fig:shg300K} for comparison).  The SHG intensity $I_{2\omega}$ should in principle obey the relationship $I_{2\omega} \propto S^4$, where $S$ is the magnitude of the sublattice magnetization \cite{SHGinNiO}.  Based on this one could infer that the magnetic order in the excited state drops on the order of several percent. 
To better understand this apparent discrepancy we need to examine more closely what precisely the magnetic diffraction signal measures.

From Eq.\ \ref{eq:M} we see that the magnetic scattering factor is related to a sum of projections of  $\mathbf{S}(\mathbf{Q})$ and $\mathbf{L}(\mathbf{Q})$, Fourier components of the spin density and an orbital density function.  For our  measurement geometry and x-ray polarizations the scattering factor is given by 
\begin{equation}
M_{\pi\sigma} = 2\sin^2{\rm{\theta}} \left[ \cos{\rm{\theta}} \left(L_1 + S_1 \right) + \sin{\rm{\theta}} S_3\right].
\label{eq:Mps}	
\end{equation}
Here, $L_1$ and $S_1$ are the components of $\mathbf{L}(\mathbf{Q})$ and $\mathbf{S}(\mathbf{Q})$ along the $( 11\overline{2})$ direction as defined by Blume and Gibbs \cite{blumegibbs} and depicted in 
Fig.~\ref{fig:NiO_xrayAB}(b). $S_3$ is the component of $\mathbf{S}(\mathbf{Q})$ in the $(111)$  direction. 
We will neglect contributions from $L_1$, although it was found to lead to small contributions to the equilibrium magnetic moment in NiO \cite{bergevinNiO}. The diffracted intensity is then
\begin{equation}
I(\rm{\theta})_{\pi\sigma} \propto \sin^2{\rm{\theta}} \tan{\rm{\theta}} \left( S_1 \cos{\rm{\theta}} + S_3 \sin{\rm{\theta}}\right)^2.							
\label{eq:Ips}
\end{equation}
As discussed in the previous section, the equilibrium sublattice spins can point along any of three equivalent $\{11\overline{2}\}$ directions, resulting in three possible spin domains: $(11\overline{2})$ (domain ``A''), $(1\overline{2}1)$ 
(domain ``B''), and $(\overline{2}11)$ (domain ``C'').   Using superscripts to denote the different spin domains with their respective ratios of the total population $a$, satisfying $a^{A} + a^{B} + a^{C} = 1$, we have in equilibrium $S_1^A = a^{A}S$, $S_1^{B,\,C} = -a^{B,\,C}S/2$, and $S_3^{A,\,B,\,C} = 0$.

The dynamics inferred from previous interpretations of the excited state imply both a reduction of the average spin moment and a reorientation of the spin sublattice vector  along $(111)$, which in our treatment would lead to a decrease in the $S_1$ component and an increase in the $S_3$ component. We can parameterize this change through new time-dependent variables $\alpha$ and $\gamma$, with $\alpha$ representing the dimensionless magnitude of the average sublattice spin and $\gamma$ the reorientation toward the $(111)$ direction.   We then have $S_1^A = \alpha \,a^{A} S \cos\gamma$, $S_1^{B,\,C} = - \alpha \,a^{B,\,C} S \cos\gamma/2$ and $S_3^{A,\,B,\,C} = \alpha \,a^{A,\,B,\,C} S \sin \gamma$.  This leads to relative changes in the diffraction intensity for each domain type
\begin{equation}
\left( {I(\rm{\theta})_{\pi\sigma} \over I(\rm{\theta})^{(0)}_{\pi\sigma} }\right)_A = \alpha^2 \left( \cos \gamma + \tan \theta \sin \gamma \right)^2
\end{equation}
\begin{equation}
\left( {I(\rm{\theta})_{\pi\sigma} \over I(\rm{\theta})^{(0)}_{\pi\sigma} }\right)_{B,\,C} = \alpha^2 \left({\frac{1}{2} \, \cos \gamma} - \tan \theta \sin \gamma \right)^2
\end{equation}
In equilibrium, $\alpha = 1$ and $\gamma = 0$. In case of spin reorientation $\gamma$ should assume some non-zero value.  In our data we observe no changes at any time within our estimated 2\% uncertainty. Since the spin domain population for the probed volume is a priori unknown, it is difficult to place a definite upper bound for changes in $\alpha$ and $\gamma$.  We can, however, point out that the x-ray data show no evidence of strong oscillatory behavior, suggesting that a re-examination of the proposed dynamics is warranted.  

\section{Time-resolved SHG measurements}

SHG has been demonstrated as a powerful tool to study ferroelectricity and antiferromagnetism, where the order parameter is correlated to a breaking of the inversion symmetry. It is, however, remarkable that the method was proven to be applicable also to centrosymmetric materials such as NiO and CoO when the generation is resonantly enhanced and a magnetic dipole transition participates in the excitation \cite{SHGinNiO}. In NiO this means, however, that SHG is only efficient when using probe light of a narrow band around 1200 nm. Even then, the generated second harmonic intensity is very weak, requiring intense probe light pulses and long acquisition times. 
All previous studies were therefore carried out using a fundamental wavelength $\lambda =$ 1200 nm and detected the SHG that was emitted in an anti-collinear geometry, leading to the observation of oscillations in the SHG response with a frequency of either 55 GHz or 1 GHz\cite{SHGpumpprobe, SHGpumpprobe2, SHGpulselength}. 

Here, we extend these SHG studies with time-resolved measurements in three major respects in an effort to understand these dynamics more completely.  First, we carried out measurements in both reflection and  transmission geometries to test the sensitivity of the dynamics to the phase-matching conditions of the SHG process.  Second, we performed measurements at multiple probe wavelengths to determine whether the observed oscillations change with probe frequency.  Third, we measured the dynamics with different sample thicknesses to determine any possible role of the distance between the interfaces.  We also re-investigated the pump intensity dependence of the dynamics.  
Our results show that the observed dynamics depend strongly on probe wavelength, geometry and acoustic properties of the samples and appear to be consistent with simulations of the time-dependent SHG process when allowing for a strain-induced perturbation of the linear and non-linear susceptibilities.

For these experiments, we employed an amplified 800 nm Ti-sapphire laser system providing 100 fs pulses at 1 kHz repetition rate. A fraction of the output beam was fed into an optical-parametric-amplifier that generated probe pulses tunable between 1140 and 1400 nm wavelength. To allow for the investigation of dynamics extending over several nanoseconds,  the probe beam made two round trips over a 1.3 m long delay stage. Pump and probe beams were combined at a relative angle of 2$^{\circ}$ using a dichroic mirror. The 800 nm pump and IR probe beam were then delivered to the sample with typical pulse energies of 120 \textmu J and 20 \textmu J, respectively, and filtered for second harmonic light generated by the optical components. After focusing this leads to considerable pump and probe peak fluences of 50 and 240 mJ/cm$^2$. 
After the sample, a series of polarizers and color filters filtered out both the fundamental probe and scattered 800 nm light, as well as third harmonic light and multi-photon fluorescence. The SHG was detected using a GaAs photomultiplier tube. 

\subsection{Reflection geometry measurements}

\begin{figure}[tbp]
	\centering
	\includegraphics[width=.5\textwidth]{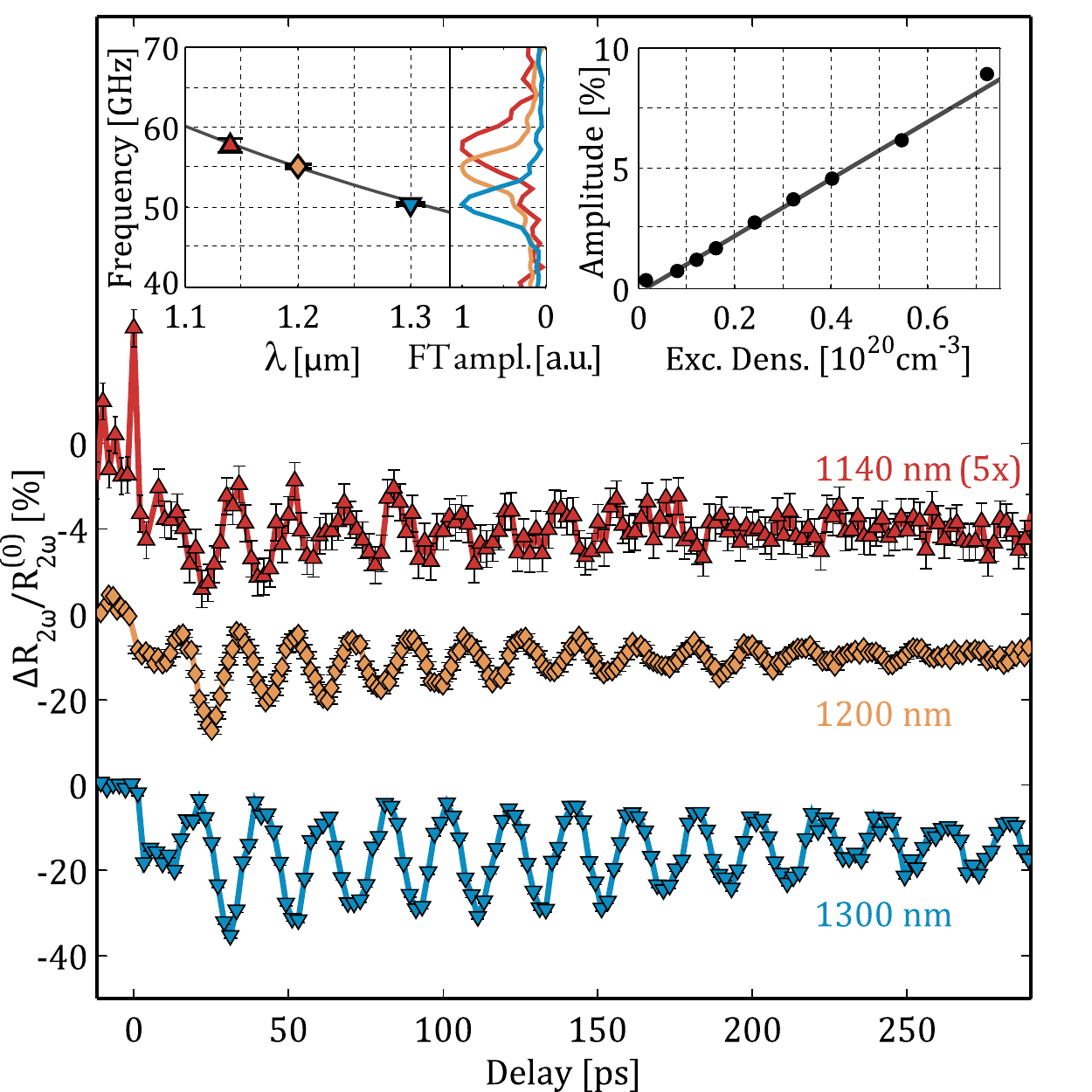}
	\caption{Relative change in SHG in reflection geometry for 3 different wavelengths in the 36 \textmu m thick sample. 
Lines are shown to guide the eye and the measurement of $\lambda = 1140$ nm is scaled for better comparison. Its smaller modulation amplitude is mainly due to an increase in fluorescence background for shorter wavelengths. The Fourier transforms and their corresponding peaks are shown in the left inset, together with the result for $\omega_{+}$ given by Eq.\ \ref{eq:Tfreqsimple}. The right inset shows the dependence of the oscillatory amplitude on excitation density at $\lambda = 1200$ nm (error bars are smaller than marker sizes).}
	\label{fig:shg300K}
\end{figure}

Using a probe beam wavelength of 1200 nm and detecting the SHG light in reflection geometry, we were able to reproduce the SHG response after optical excitation as previously reported \cite{SHGpumpprobe} in a 36 \textmu m thick sample that shows evenly distributed small domains, as well as in a 50 \textmu m thick sample showing domain sizes of around 100 \textmu m, see Fig.\ \ref{fig:shg300K}. The SHG response at 1200 nm features a 55 GHz oscillation, showing an amplitude of about 10\% and a decay time of above 250 ps. A careful study of the observed oscillation amplitude on fluence and excitation pulse length did, however, not show any deviation from linear behavior, as shown in the right inset in Fig.\ \ref{fig:shg300K}. This stands in contrast to earlier publications that observed a threshold behavior depending on pulse peak power \cite{SHGpumpprobe, SHGpumpprobe2, SHGpulselength}. 

\begin{figure*}[tb]
	\centering
	\includegraphics[width = .925\textwidth]{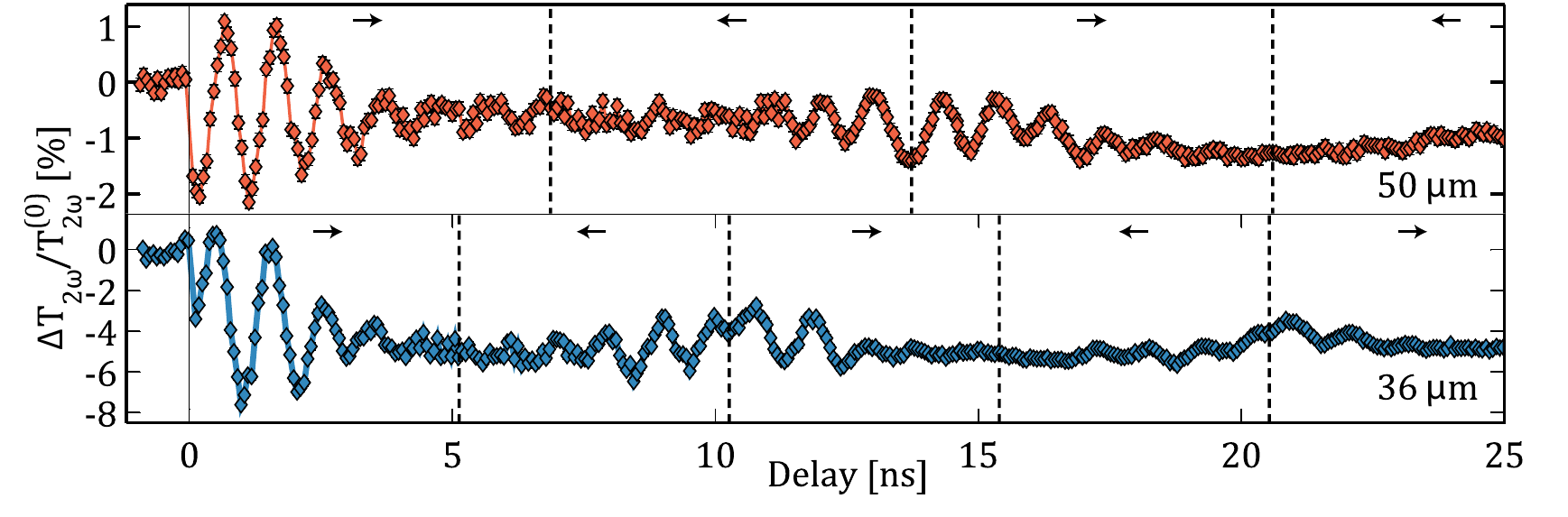}
	\caption{SHG response to optical excitation in collinear geometry acquired at 1200 nm. The oscillations show a recovery in amplitude centered at 10.3 and 13.7 ns for the 36 and 50 \textmu m sample, respectively. This is coincident with the observation of phase discontinuities at these times, that also occur at 5.15 and 6.85 ns. Pump probe traces show similarity to every other time segment, while consecutive segments appear as mirrored (indicated by arrows).}
	\label{fig:shgcollong}
\end{figure*}

A remarkable observation can be made when detuning the probe wavelength away from the resonance condition at 1200 nm. This is experimentally challenging, as the SHG intensity is largely reduced, which prevented such detuning in earlier experiments. Nonetheless, over an accessible wavelength range between $\lambda = 1140$ and $1300$ nm, we were able to acquire data showing a dependence of the observed oscillations on probe wavelength. This suggests that the frequencies of the oscillations are not an intrinsic property of the material, but arise instead from an interaction between material and the probe beam. These measurements were carried out at room temperature and repeated at 150 K to test for temperature dependencies. Within the precision of our measurement, we did not, however, observe significant differences in the measured frequencies. 

In addition, we observed considerable differences in the SHG yield for different NiO samples. SHG from a 
1 mm thick bulk crystal was undetectable, while 36 \textmu m and 50 \textmu m thick samples gave observable yield in reflection geometry. As optical properties and the domain structure in NiO vary depending on their exact stoichiometry \cite{NiOoptprop}, this might be caused by differences in the annealing process. It seems, however, more probable that these differences reflect instead a thickness dependence, as the slab thicknesses are of the same order as the $\omega$ and $2\omega$ absorption lengths of 23 \textmu m and 35 \textmu m, respectively. This would imply that the detected signal in reflection geometry consists largely of internally reflected SHG. For this reason, we also studied collinearly generated SHG in a transmission geometry.

\subsection{Transmission geometry measurements}

Figure \ref{fig:shgcollong} shows the SHG response acquired in transmission geometry over a time interval of 25 ns after excitation for two different slab thicknesses. Using 1200 nm as probe wavelength, the SHG shows pronounced oscillations with a frequency of about 1 GHz, as reported previously for some measurements performed in a reflection geometry \cite{SHGpulselength}. No signs of oscillations in the 50 GHz range could be observed. The SHG yield in this configuration is in fact about two orders of magnitude larger than in the reflection geometry. 
As the refractive index for the second harmonic $n_{2\omega} \approx 2.4$ implies a Fresnel reflectivity of 17\% and the absorption length is comparable to the sample thickness for the frequencies considered here, this implies that the major part of observed SHG in reflection geometry is in fact due to internal reflection of the collinearly created second harmonic light. 
Moreover, the data shown in Fig.\ \ref{fig:shgcollong} feature recurrences of the envelope amplitude as well as phase discontinuities that are especially remarkable in the 36 \textmu m thick sample. The observation of these phase discontinuities suggests that the recovery of amplitudes is not caused by a beating\cite{SHGpulselength} but rather by a reflection.

There is also a remarkable correlation between the crystal thickness and the observed recurrence times. The times at which the phase discontinuities occur corresponds to integer multiples of the acoustic round trip time $T_{ac} = 2d / v_{s}$ when using a speed of sound $v_{s} = (7.1\pm 0.1)$ km/s in good agreement with ultrasound measurements \cite{elastic}.

\begin{figure}[htbp]
	\centering
	\includegraphics[width=0.5\textwidth]{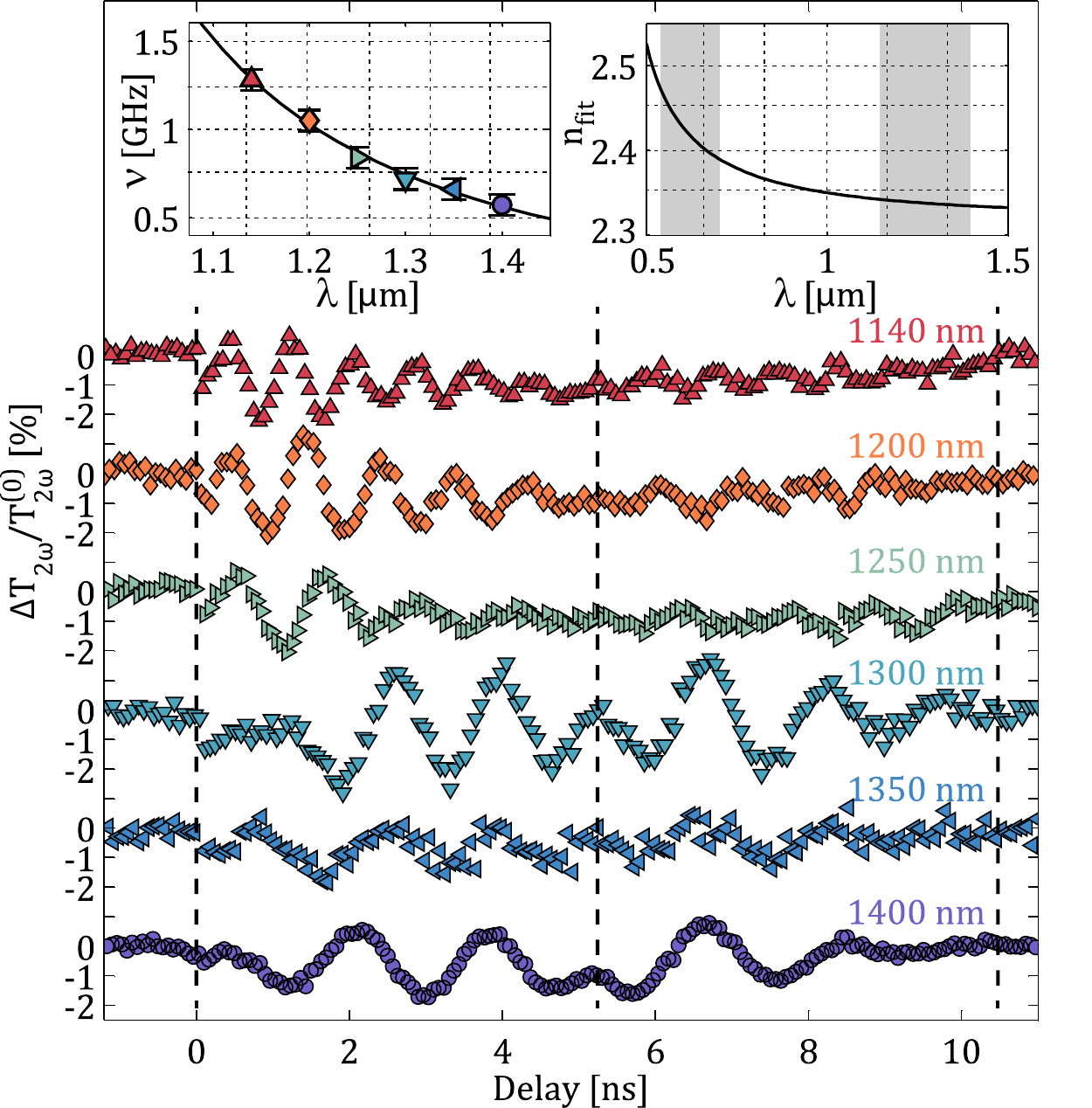}
	\caption{SHG response to optical excitation in transmission geometry for various probe wavelengths in the accessible wavelength range. A common feature of all traces is a time inversion symmetry with respect to $t = 5.15$ ns  indicated by a broken line. A strong frequency dependence of the induced oscillations is observed. The maximum positions of their Fourier transforms over the first 5.15 ns is shown in the inset on the left, which also shows the resulting frequencies based on the simulation and following Eq.\ \ref{eq:Tfreqsimple}. The fit result of the real part of the refractive index is shown to the right. Relevant areas are highlighted.}
	\label{fig:shgcol}
\end{figure}

In Fig.\ \ref{fig:shgcol} we also investigated the dependence of the observed dynamics on probe wavelength in a transmission geometry in the 36 \textmu m thick sample.
A pronounced frequency and phase dependence can be observed over a probe wavelength range between 1140 and 1400 nm. The left inset in Fig.\ \ref{fig:shgcol} shows frequencies estimated from the Fourier transform based on the first 5.15 ns after excitation. Phase discontinuities appear for each wavelength at 5.15 ns, after which the dynamics display mirror-symmetric behavior that manifests as an apparent reversal of the direction of the time axis. 

\section{Modelling the SHG response to optical excitation of a spatially confined semi-transparent crystal}

For optically absorbing materials, Thomsen et al.\ presented a model to describe how light-induced strain is driven by the thermal stress from a sudden increase in temperature after optical excitation \cite{StrainWave}. We will briefly review these findings in the following. Due to the absorption profile in the crystal and the boundary conditions introduced by the surfaces, where applied stress is compensated, the induced stress is strongly inhomogeneous in the direction normal to the surface. 
With density $\rho$, strain $\eta_{33}$, stress $\sigma_{33}$ and the displacement in z direction $u_{3}$, (here $\left[111\right]$), the equations of elasticity can be expressed as \cite{StrainWave} 
\begin{equation}
	\rho \frac{\partial^2 u_3}{\partial t^2} =\frac{\partial\sigma_{33}}{\partial z}, \,\,\,\,\,	\eta_{33} = \frac{\partial u_{3}}{\partial z}.
\end{equation}
With the boundary conditions given by the crystal surfaces and the initial condition of zero strain before excitation, we obtain solutions 
\begin{equation}
	\eta_{33}(z,t) = (1-R) \frac{\alpha Q}{A}\frac{b}{C_V}\frac{1+\nu}{1-\nu} F(z, t),
	\label{eq:strainsol}
\end{equation}
where (1$-$$R$) accounts for the reflection loss, with the 800 nm reflectivity $R$ and $\alpha Q/A$ gives the excitation density, with $\alpha$ being the 800 nm intensity absorption coefficient, $A$ the excited sample area and pulse energy $Q$. $\frac{b}{C_V}\frac{1+\nu}{1-\nu}$ is the material specific geometric response to heat with the expansion coefficient $b$, specific heat capacity $C_{V}$ and Poisson's ratio $\nu$. $F(z, t)$ is the dimensionless solution to the equation of elasticity based on the given boundary conditions, given in appendix \ref{appendixa}. In our model, an exponential decay term $e^{-t/t_d}$ is multiplied with Eq.\ \ref{eq:strainsol} to approximate the effects of acoustic loss and heat dissipation. 
Since NiO is semi-transparent over the range of relevant wavelengths,
it is necessary to consider the whole volume of the thin slabs, including strain waves propagating from both interfaces. The parameters employed for the calculation are summarized in Table \ref{tab:ElasticAndThermalParameters}.

As strain implies local variations of interatomic distances, it affects not only the mechanical but also the optical properties of the crystal. Typical deformations correspond to a relative length change of $10^{-4}$, hence the influence on optical parameters, such as the permittivity $\epsilon$, can be treated as perturbations
\begin{equation}
	\Delta \epsilon(z,t) = 2(n+i\kappa)\left[\frac{\partial n}{\partial \eta_{33}} + i \frac{\partial \kappa}{\partial \eta_{33}}\right]\eta_{33}(z,t),
	\label{eq:dndeta}
\end{equation}
where $n$ and $\kappa$ are the real and imaginary parts of the refractive index.  Here we also assume that the temperature changes have no direct effect on $\epsilon$. 
Values of $\partial n/\partial \eta_{33} \ne 0$ have previously been observed in NiO films as well as in $(001)$-cut bulk NiO using reflectivity and ellipticity measurements \cite{BoscoNiO, StrainNiO}. A modulation of the linear refractive index already implies possible consequences for the SHG response arising from changes in the SHG coherence length. 

In a similar manner, we can also parameterize possible changes in the second-order susceptibity $\chi^{(2)}$ with respect to changes in temperature and strain
\begin{equation}
\begin{split}
\Delta\chi^{(2)}(z,t) = \frac{\partial\chi^{(2)}}{\partial T}\bigg|_{T = \rm{RT}} \Delta T (1-e^{-t/t_{m}}) e^{-t/t_{d}} \\
 + \frac{\partial \chi^{(2)}}{\partial \eta_{33}}\bigg|_{\eta_{33} = 0} \eta_{33}(z,t).
\end{split}
\label{eq:dchi2deta}
\end{equation}
where RT stands for room temperature, $t_m$ denotes the demagnetization time, which is in AFM compounds typically few ps \cite{demagCr2O3}, and $t_d$ is the diffusion time constant $t_d$ which lies in the ns regime. 
For the sake of simplicity of our model, we do not take into account the imaginary component of the nonlinear refractive index.

The second order susceptibility  $\chi^{(2)}$ should depend on the antiferromagnetic order parameter $l$, which is related to magnetoelastic lattice distortions \cite{SHGinNiO}. This in turn depends on the temperature, which gives us an expression for the temperature dependence 
\begin{equation}
\chi^{(2)}(T) \propto (1 - T/T_N)^{2\beta},
\label{eq:lsquare}
\end{equation}
with critical exponent $\beta = 0.33$ \cite{criticalexp}. The local temperature changes due to optical excitation in our experiment are on the order of several K at room temperature, justifying a linear expansion in $\Delta T$.

A dependence of $\chi^{(2)}$ on strain can arise from at least two physical origins.  One possibility is a strain-induced change in the energies of the $(3d)^8$ states of Ni$^{2+}$ that may alter the double-resonance condition of SHG in NiO.  Another possibility is  magnetoelastic effects that more directly change the sublattice magnetization $l$.  Both of these possibilities are at present beyond our abilities to quantify, and so we simply incorporate $\partial\chi^{(2)}/ \partial \eta_{33}$ 
as a parameter in our model.

For a full quantitative simulation of the time dependent SHG in NiO, we apply a time-resolved propagation matrix calculation. This approach is  suitable for our experimental geometries as internal reflections are not negligible. To implement temporal and spatial variations, the crystal volume is divided along the longitudinal direction in $N$ slices of each a few nanometer thickness (well below the scale of optical wavelengths) whose optical properties are given by Eq.\ \ref{eq:dndeta} and \ref{eq:dchi2deta}. Restricting the problem to one dimension is permitted as the Rayleigh range is many times larger than the crystal thickness. Each slice with index $m$ can then be treated as a source of SHG $S_m$ due to the induced nonlinear polarization $P^{NL}_m$
\begin{equation}
\begin{split}
S_m =& -\mu_0 \frac{\partial^2}{\partial t^2} P^{NL}_m, \\
P^{NL}_m =& -i \chi^{(2)}_m(t) (E^{+}_{\omega, m}(t) + E^{-}_{\omega, m}(t))^2,
\end{split}
\label{eq:source}
\end{equation}
where $\mu_0$ is the vacuum permeability and $\chi^{(2)}_m$ represents the magnetic dipole assisted nonlinear susceptibility according to Eq.\ \ref{eq:dchi2deta}. $E^{+/-}_{\omega, m}$ correspond to the right- and left-ward propagating fundamental fields at slice $m$. 
The small value of the second order susceptibility $\chi^{(2)}$ allows us to describe the SHG process in the Born approximation in which the fundamental fields $E_{\omega}$ are independent of the SHG. We can then apply the propagation matrix method to derive $E^{+/-}_{\omega, m}$ at each slice position and calculate the emitted second harmonic fields as described in appendix \ref{appendixb}.

\begin{figure}[htbp]
	\centering
	\includegraphics[width=0.50\textwidth]{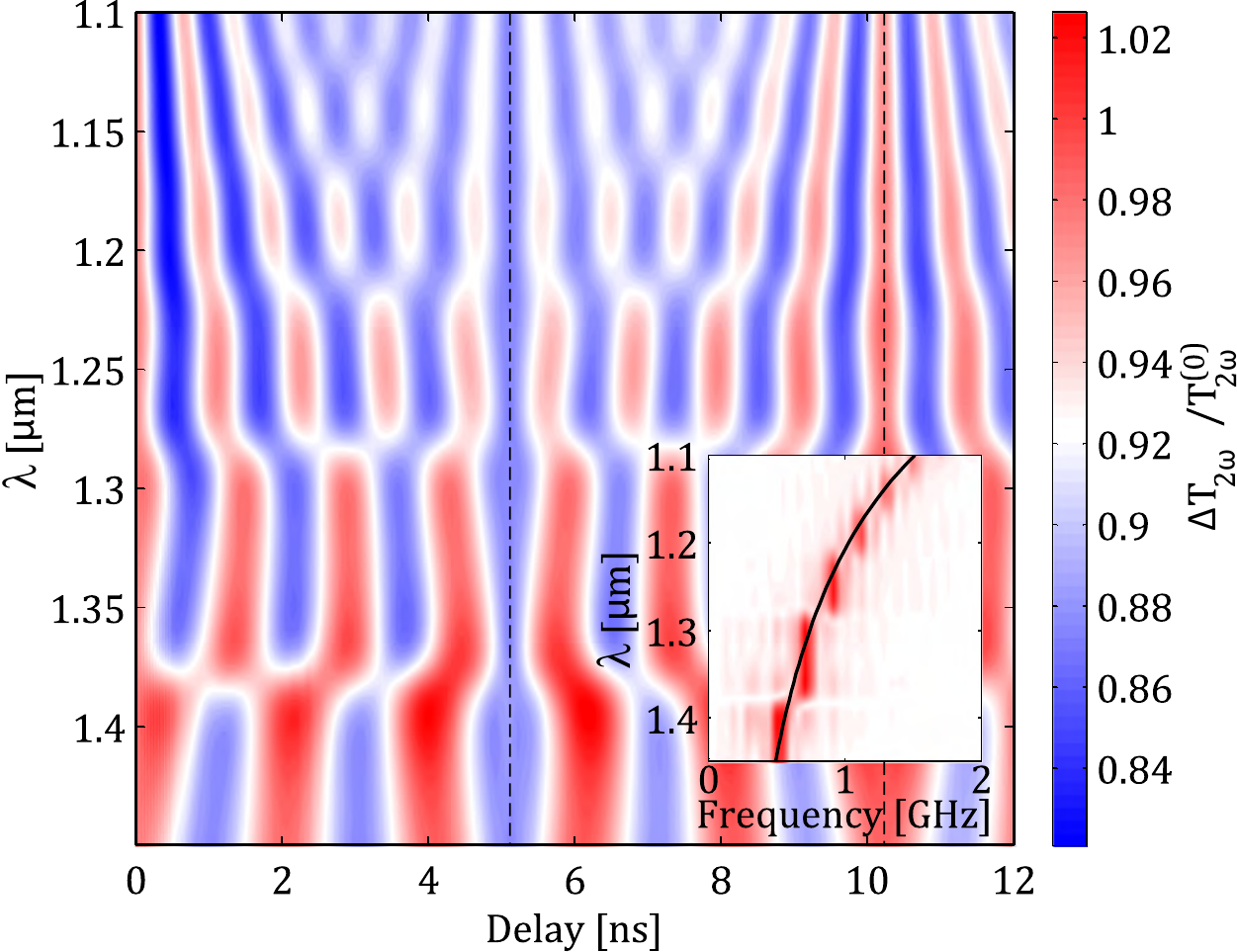}
	\caption{Simulation of the relative change in \textit{transmitted} SHG in the presence of optically induced propagating strain. The calculation solves the propagation matrix model for a fluence of 50 mJ/cm$^2$ and a NiO crystal of 36 \textmu m thickness. Acoustic reflections occur at 5.15 ns and are marked as broken lines. The Fourier transform is shown in the inset where the solid line shows the curve given by $\omega_{-}$ in Eq.\ \ref{eq:Tfreqsimple}. The spectrum shows step-like discontinuities due to the acoustic reflections. A direct comparison to the data is shown in Fig.\ \ref{fig:shgcol}.}
	\label{fig:modelT}
\end{figure}

\begin{figure}[H]
	\centering
	\includegraphics[width=0.50\textwidth]{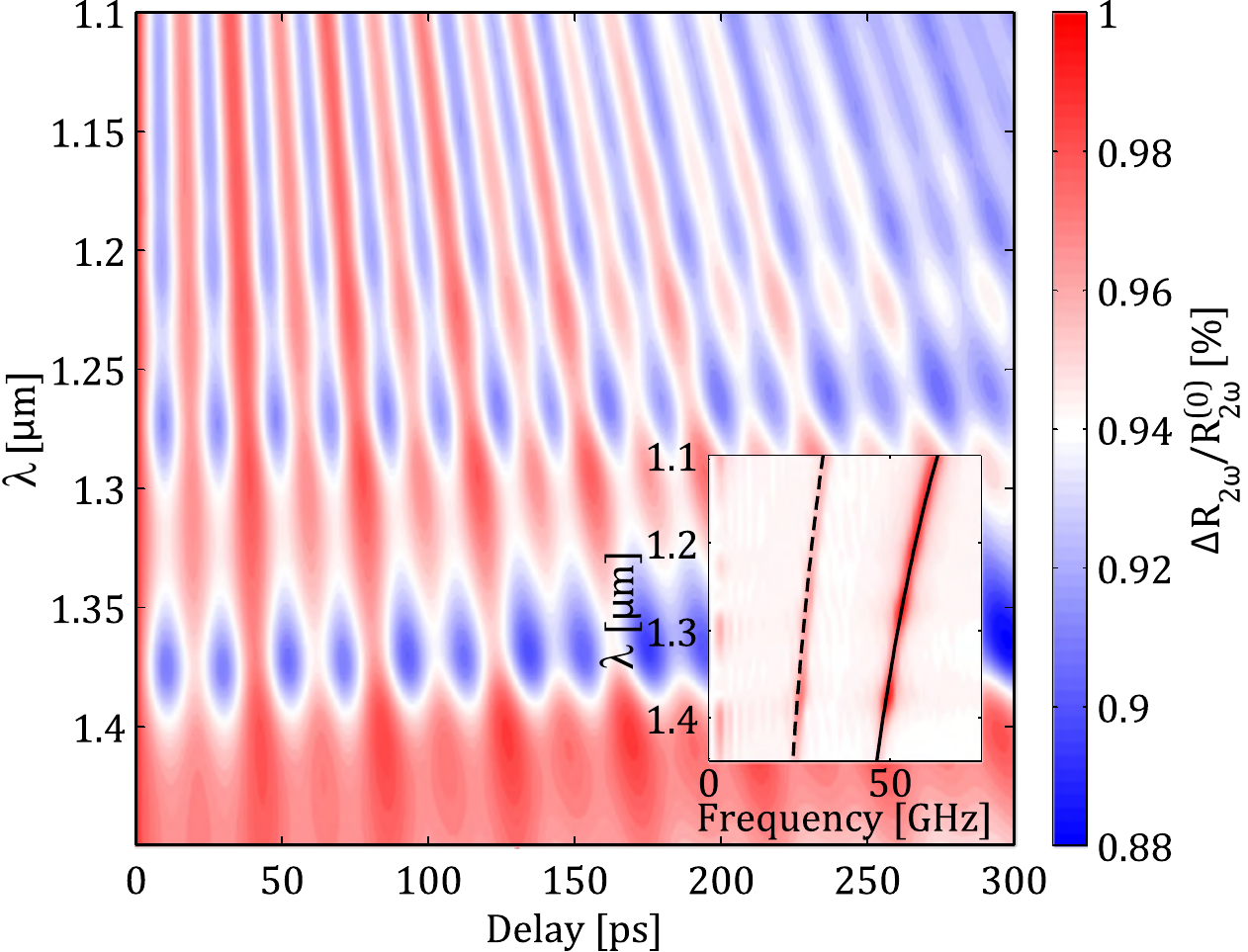}
	\caption{Simulation of the relative change in \textit{reflected} SHG in the presence of optically induced propagating strain. The calculation shown here uses the same parameters as the simulation in Fig.\ \ref{fig:modelT}. The spectrum is shown in the inset, together with the frequency of Brillouin scattering of the fundamental $\omega_B = 2 n v_s \omega/c$  and the curve given by $\omega_{+}$ of Eq.\ \ref{eq:Tfreqsimple}.  These are shown as broken and solid lines, respectively. The simulation, in particular for longer wavelengths, contains spectral components at $\omega_B$. This is not reflected by our measurements and most likely due to an underestimation of absorption and overestimation of $\partial n_{\omega}/ \partial \eta_{33}$ at the fundamental wavelength. However, small contributions were previously observed at $\lambda = 1200$ nm \cite{SHGpumpprobe}.}
\label{fig:modelR}
\end{figure}

Results of the simulation for transmitted and reflected intensities are shown in Fig.\ \ref{fig:modelT} and \ref{fig:modelR}, respectively. On ns timescales, the SHG in reflection geometry follows the behavior of the transmitted SHG, which reflects the fact that its main source is the collinear generation process and internal reflection, with additional contributions of the SHG arising from internally reflected fundamental light. The real part of the refractive index $n(\omega)$ used in our simulations was obtained from a fit to the experimental data as will be discussed below. Further optical parameters employed in these calculations are given in appendix \ref{appendixc} and Table \ref{tab:optparameters}.


\section{Discussion of the optical results}

According to our model, the dynamics in SHG are the result of the time dependent second order polarization and light propagation inside the medium in the presence of coherent acoustic waves. By using the propagation matrix method the model takes into account multiple reflections of fundamental and second harmonic fields as well as the strain waves inside the crystal. This is especially important as a major part of the SHG in the reflection geometry arises from the internal reflection from the interface on the opposite side of the crystal. \\

Although to obtain approximate quantitative accuracy the model includes many different effects, it is possible to gain some additional insight by considering in general the effect of strain waves on phase matching for a collinear or anticollinear SHG process.
In a uniform, non-absorbing medium, the output intensity in the first Born approximation is related to the effective length $L$ by
\begin{equation}
\begin{split}
I_{2\omega} &\propto \left| \int_0^L \chi^{(2)} \exp(i \Delta k z) dz\right|^2 \\&= \left|{\chi^{(2)} \over \Delta k}\right|^2 4 \sin^2 (\Delta k L /2)
\end{split}
\end{equation}
where $\Delta k = 2 k_{\omega} \pm k_{2\omega}$, with the sign depending on whether the fundamental and SHG beams are collinear or anti-collinear. 
Strain-induced modulation of the linear and nonlinear optical coefficients causes small, z-dependent changes in both $\Delta k$ and $\chi^{(2)}$.  If we approximate a strain wave propagating through the crystal away from the front interface as a real-valued discontinuity in both $\Delta k$ and $\chi^{(2)}$ that moves with the sound velocity $v_s$, we obtain
\begin{equation}
\begin{split}
I_{2\omega} \propto &\bigg| \int_0^{v_st} (\chi^{(2)} + \delta_{\chi^{(2)}}) \exp(i (\Delta k + \delta_{\Delta k}) z) dz  \\&+ \int_{v_st}^L \chi^{(2)} \exp(i \Delta k z) dz\bigg|^2 
\end{split}
\end{equation}
where we assume that beyond the discontinuity the optical constants are unperturbed, and before the discontinuity the second-order susceptibility changes by $\delta_{\chi^{(2)}}$ and the phase mismatch $\Delta k$ changes by $\delta_{\Delta k}$.  Evaluating this to first order in $\delta_{\chi^{(2)}}$ and $\delta_{\Delta k}$ yields
\begin{equation}
\begin{split}
I_{2\omega} &\propto \left|{\chi^{(2)} \over \Delta k}\right|^2 \bigg[4 \sin^2 (\Delta k L /2) \\&+ 2 \delta_{\Delta k} v_s t (\sin (\Delta k (L-v_st) + \sin (\Delta k v_s t)) \\ &+ 2 \left( {\delta_{\chi^{(2)}} \over \chi^{(2)}} - {\delta_{\Delta k} \over \Delta k} \right) \Big( 2 \sin^2 (\Delta k L /2) \\&+ \cos (\Delta k (L-v_st)) - \cos (\Delta k v_st) \Big)\bigg].
\label{eq:simplediscon}
\end{split}
\end{equation}
We see immediately from this expression that the intensity of the SHG is modulated in time with a frequency
\begin{equation}
\omega_{+/-} = \operatorname{Re}(2 k_{\omega} \pm k_{2\omega}) v_{s}.
\label{eq:Tfreqsimple}
\end{equation}
where we explicitly take the real part of the expression for $\Delta k$ to extend our result to the more realistic case where there is a small imaginary component to both $k_\omega$ and $k_{2 \omega}$.

The $\omega_-$ values from Eq.\ \ref{eq:Tfreqsimple} correspond to the slow oscillations in the transmission geometry at frequencies near 1 GHz as shown in Fig.\ \ref{fig:shgcollong} and \ref{fig:shgcol}, where we show values for $\omega_-$ at various probe frequencies by extracting values of the refractive index from a four-parameter Sellmeier equation. The result for $n(\lambda)$ is shown in the right inset in Fig.\ \ref{fig:shgcol} and the parameters are given in the appendix in Table \ref{tab:optparameters}. 
This parameterization is only a coarse approximation of the linear optical dispersion, since the absorption spectrum of NiO shows strong features in the observed spectral range that are not accounted for in this model \cite{NiOoptprop}. 
Both the simulations and Eq.\ \ref{eq:simplediscon} predict that the onset phase of these oscillations should depend on the unperturbed value of $\Delta k$, which in turn depends sensitively on the probe wavelength.  
Different sample thicknesses $L$ also lead to different onset-phases for the same probe wavelength, as observed in Fig.\ \ref{fig:shgcollong}. The simulation can also account for the observed temporary increase in absolute SHG above the equilibrium level, since the strain modulation is under some circumstances able to effectively improve phase matching for higher outcoupling.

The $\omega_+$ values from Eq.\ \ref{eq:Tfreqsimple} correspond to the fast oscillations observed in the reflection geometry.  These values are plotted in the inset of Fig.\ \ref{fig:modelR} and match the measured data quite well.

In order to match the magnitude of the oscillations seen in experiment,  the simulations require a large magnitude of $\partial \chi^{(2)}/\partial \eta_{33}$, leading to local changes in $\chi^{(2)}(z)$ of several percent (see Table \ref{tab:optparameters}). 
The physical origin behind this coupling is unclear, but may be found in the same magnetoelastic interaction responsible for the large temperature dependence of $\chi^{(2)}$\cite{SHGinNiO}. In principle, our observations could also be brought into agreement with a magnetoelastic or flexoelectric contribution to SHG due to $\partial \chi^{(2)}/\partial(\frac{d \eta}{d z}) \ne 0$ or higher orders \cite{NiOstrainreorient}. The exact dependence on strain, however, is outside the scope of this work as the applied experimental techniques do not allow us to distinguish between different possible coupling mechanisms. Further insight could be obtained using static SHG imaging of strained crystals or by employing a strongly focused probe beam in time resolved SHG in order to disentangle contributions arising from different depths inside the crystal.

The agreement between simulations and experimental data is not exact but they reproduce the observed frequencies, lead to effects of similar magnitude and give a quantitative explanation for the phase behavior of the oscillations. The relative magnitude of the dynamics are in general somewhat underestimated. 
We found the simulation results to be highly sensitive to small changes in the assumed linear optical properties. This sensitivity may account for some of these discrepancies.  
Furthermore, the large probe fluences used in these experiments far exceed the limit of small perturbation and will lead to back-action on internally reflected beams which is not taken into account. \\

As an alternative explanation for the observed effects, in principle a strain-induced modulation of $n$ also leads to coherent Brillouin scattering of the second harmonic light according to $\omega_{osc} = 2 k_{2\omega} v_s$. For wavelengths closer to the band-gap, the strain dependence of the refractive index $\partial n/\partial \eta$ strongly increases \cite{BoscoNiO}, which in this case could lead to a pronounced visibility of Brillouin scattering at around 600 nm as compared to the fundamental light at 1200 nm. 
In the simulation, both $\partial n/\partial \eta \ne 0$ and $\partial \chi^{(2)}/\partial \eta \ne 0$ by themselves can lead to modulations at the observed frequencies and also to a small spectral contribution at the Brillouin scattering frequency of the fundamental beam.
However, using our model, linear Brillouin scattering alone can not explain the large magnitude of the observed modulation in SHG of up to 30\% (see Fig.\ \ref{fig:shg300K}) as it would also lead to similarly strong modulations of the fundamental light, which was not observed. 

As a final remark it might be surprising that despite the seemingly general nature of a strain induced change of the nonlinear susceptibility, the dynamic effects observed in NiO were not found in similar compounds such as CoO and KNiF$_3$ which also rely on magnetic dipole assisted SHG. This may reflect a difference in the magnetoelastic interaction in these systems. We note, however, that in NiO  these dynamics are strongly dependent on sample thickness, absorption and dispersion in $\Delta k$.  It may be that only a narrow set of experimental parameters leads to similar dynamics. As the SHG process in these compounds is highly restricted by resonance conditions, the range of these parameters is quite limited.

\section{Conclusion}

Picosecond-time-resolved non-resonant magnetic x-ray diffraction was employed as a tool to study sublattice magnetism and suggests that the dynamics in NiO observed with SHG may not directly reflect dynamics of the antiferromagnetic order parameter. The low diffraction efficiency hindered a more precise measurement of the dynamic change of the structure factor but the result excludes a large drop in the spin sublattice magnitude as suggested by previous models.

The extended SHG data presented here supplement previous investigations and reveal some aspects that appear inconsistent with previous explanations for the ultrafast dynamics of SHG in NiO. The threshold behavior in the presence and frequency of the SHG oscillation, which was one of the main arguments for an interpretation in terms of dynamics of the order parameter \cite{SHGpumpprobe, SHGpumpprobe2, SHGpulselength}, was not verified by the present experiments. Our SHG studies give evidence for an acoustic origin of the dynamics, which becomes particularly apparent in the observation of echoes that depend on the acoustic path-length, as well as the probe wavelength dependencies for the two observed frequency regimes.

The choice of crystal dimensions used in the current, as well as in previous SHG studies on NiO requires careful interpretation of the observations as the semi-transparency allows for multiple reflections. In particular, a direct connection between ultrafast induced changes in $\chi^{(2)}$ and $l$ cannot readily be made. Implementing a strain dependence of the linear and nonlinear refractive index in a full calculation of the SHG process in a spatially confined, semi-transparent medium gives a self-consistent explanation of the observations in various geometries and over a wide range of timescales in which the two observed frequencies can be understood as arising from coherent acoustic perturbation that effectively projects the local collinear and anti-collinear phase mismatch in SHG into the time domain, with observed frequencies $\omega_{+,-} = \operatorname{Re}(2 k_{\omega} \pm k_{2\omega}) v_{s}$. These conclusions might be tested further by a direct measurement of the dependence of $\chi^{(2)}$ on strain along the $\left[111\right]$ direction.

\begin{acknowledgments}
Time resolved x-ray diffraction measurements were carried out under the proposal number 2012B8003 at beamline BL3EH2 of the free electron laser SACLA at the RIKEN SPring-8 Center, Harima Institute, Sayo, Hyogo 679-5148, Japan. We thank R. Follath for experimental support at SACLA and we thank J. Strempfer and S. Francoual for enabling us to perform preparative static magnetic x-ray diffraction measurements at the resonant scattering and diffraction beamline P09 at PETRA III, Deutsches Elektronen-Synchrotron DESY, Notkestrasse 85, 22607 Hamburg, Germany. We further thank C. Detlefs for providing the analyzer crystal used in the diffraction experiment.  MF thanks N.P. Duong and T. Satoh for useful discussion. We acknowledge financial support by the ETH Femtosecond and Attosecond Science and Technology (ETH-FAST) initiative of the NCCR Molecular Ultrafast Science and Technology (NCCR MUST) program, a research instrument of the Swiss National Science Foundation (SNSF). 
\end{acknowledgments}

\appendix
\section{Surface generated strain in a confined medium}
\label{appendixa}
Equation \ref{eq:strainsol} gives the solution for strain induced by impulsive optical excitation. Its dimensionless spatial and temporal evolution $F(z,t)$ for thin crystals is then given by 

\begin{equation}
\begin{split}
F(z, t) &= f(z) \left( 1 - \frac{1}{2} f(v_s \tau(t))\right) - \frac{1}{2} f(|z - v_s \tau(t)|) \\
					\cdot \, &\operatorname{sgn}(z - v_s(t) \tau(t)) \\
								 + &f(d) \left(\frac{1}{2} f(d - z + v_s \tau(t))- \frac{1}{2} f(|d - z - v_s \tau(t)|) \right) \\
					\cdot \, &\operatorname{sgn}(d - z - v_s \tau(t)), 			 
\end{split}
\label{eq:solution1}
\end{equation}
with the initial spatial distribution of elastic stress
\begin{equation}
f(z')  = \frac{e^{\alpha(2d -z')} + R e^{\alpha z'}}{e^{2 \alpha d} -R^2},
\label{eq:solution0}
\end{equation}
and the effective time $\tau(t)$ 
\begin{equation}
\tau(t) = 
\begin{cases}
    t \operatorname{mod}(d/v_s),  &\text{if} \,\, t \operatorname{mod}(2d/v_s) \leq d/v_s \\
    -t \operatorname{mod}(d/v_s) + d/v_s, &\text{otherwise.} 
\end{cases}
\end{equation}
This definition of $\tau(t)$ emphasizes the time inversion symmetry with respect to the acoustic reflections occurring at multiples of $d/v_s$.

\begin{table}[htbp]
	\centering
		\begin{tabular}{cccccc}
		\hline
		\hline
			 $\rho$ $[\frac{\rm{g}}{\rm{cm}^3}]$ \hspace{2mm} & B $[$GPa$]$ & $\nu$ & $b$ $[\frac{10^{-5}}{\rm{K}}]$\hspace{2mm}  & $C_V$ $[\frac{\rm{J}}{\rm{cm}^3\rm{K}}]$\hspace{2mm}  & $t_d$ $[$ns$]$ \\
			\hline
			6.81$^{[}$\cite{neutrondiff}$^{]}$ & 193.8$^{[}$ \cite{elastic}$^{]}$ & 0.26$^{[}$\cite{elastic}$^{]}$ & 4.2$^{[}$\cite{expansion}$^{]}$ & 0.59$^{[}$\cite{heatcapacity}$^{]}$ & 5.0 \\
			\hline
			\hline
		\end{tabular}
	\caption{Elastic and thermal parameters of NiO used to calculate strain in the simulation. The bulk modulus $B$ and Poisson's ratio $\nu$ are based on ultrasound measurements on a crystal in (111) surface orientation \cite{elastic}. The parameter $t_d$ gives the diffusion time constant used in the simulation.}
	\label{tab:ElasticAndThermalParameters}
\end{table}

\section{Propagation matrix based calculation of SHG}
\label{appendixb}
The Born approximation allows us to use the propagation matrix approach \cite{bornwolf} to independently solve for the fundamental fields inside the crystal
\begin{equation}
\left(\begin{array}{c} E^{+}_{\omega, M} \\ E^{-}_{\omega, M} \end{array}\right) =  \mathbf{P}_{M-1}(t) \left(\begin{array}{c} E^{+}_{\omega, 0} \\ E^{-}_{\omega, 0} \end{array}\right).
\label{eq:bril}
\end{equation}
Here, $\mathbf{P}_{M-1}$ is a $2\times2$ matrix connecting the right- and leftward propagating fields at slice 0 with the fields in slice $M$. Fields at arbitrary slice positions can be derived by solving for the transmitted and reflected fields $E^{+}_{\omega, N+1}$ and $E^{-}_{\omega, 0}$, using the boundary conditions $E^{+}_{\omega, 0} = E_0$ and $E^{-}_{\omega, N+1} = 0$.

Given that the lifetime of light inside the crystal is much shorter than $\lambda/v_s$, the propagation matrices $\mathbf{P}_M(t)$ can be calculated as
\begin{equation}
\mathbf{P}_{M}(t) = \prod_{m = M}^{0} \mathbf{p}_{m}(t),
\label{eq:propmat}
\end{equation}
where propagation through a single slice is given by
\begin{equation}
\mathbf{p}_{m}(t) = \frac{1}{1-r}\left(\begin{array}{cc} 1 & -r \\ -r & 1 \end{array} \right) \cdot \left(\begin{array}{cc} e^{-\frac{i 2 \pi\Delta}{\lambda} n_m(t)} & 0 \\ 0 & e^{\frac{i 2 \pi\Delta}{\lambda} n_m(t)} \end{array} \right),
\label{eq:propmat2}
\end{equation}
with $r = (n_{m+1}(t) - n_{m}(t))/(n_{m+1}(t) + n_{m}(t))$, at normal incidence and $n_m(t)$ corresponds to the time dependent refractive index of slice $m$, while $\Delta$ represents the chosen slice thickness. Surface boundaries are included by setting $n_0(t) = n_{N+1}(t) = 1$.

The time dependent solutions of $E^{-}_{\omega, 0}(t)$ correspond to the familiar results of coherent Brillouin scattering of the fundamental beam. By knowing the fundamental field in time and space, it is then possible to calculate the emitted second harmonic light by applying the propagation matrix approach to each slice as a source of SHG according to Eq.\ \ref{eq:source}
\begin{equation}
\begin{split}
\left(\begin{array}{c} E^{+}_{2\omega, N+1} \\ E^{-}_{2\omega, N+1} \end{array}\right) = \left[ \prod_{m = N}^{0} \mathbf{P}_{2\omega, m}(t) \right] \left(\begin{array}{c} E^{+}_{2\omega, 0} \\ E^{-}_{2\omega, 0} \end{array}\right) + \\
 \sum^{N}_{k = 1}{\left[ \prod_{m = N}^{ k+1} \mathbf{P}_{2\omega, m}(t) \right] \left(\begin{array}{c} S_{k} \\ S_{k} \end{array}\right)}.
\label{eq:SHGmat}
\end{split}
\end{equation}
Equation \ref{eq:SHGmat} can be solved for the emitted SHG fields $E^{+}_{2\omega, N+1}$ and $E^{-}_{2\omega, 0}$ using source terms according to Eq.\ \ref{eq:source}, as well as the boundary condition $E^{-}_{2\omega, N+1} = E^{+}_{2\omega, 0} = 0$. \\
With the definitions
\begin{equation}
\begin{split}
\left(\begin{array}{cc} A & B \\ C & D \end{array}\right) =&  \prod_{m = N}^{0} \mathbf{P}_{2\omega, m}(t), \\
\left(\begin{array}{c} S^{+} \\ S^{-} \end{array}\right) =&  \sum^{N}_{k = 1}{\left[ \prod_{m = N}^{ k+1} \mathbf{P}_{2\omega, m}(t) \right] \left(\begin{array}{c} S_{k} \\ S_{k} \end{array}\right)}, 
\end{split}
\label{eq:sourcedef}
\end{equation}
the solutions for SHG in transmission and reflection geometry are given by 
\begin{equation}
\left(\begin{array}{c} E^{+}_{2\omega, N+1} \\ E^{-}_{2\omega, 0} \end{array}\right) =  \left(\begin{array}{c} S^{+} -\frac{B}{D} S^{-} \\ -\frac{1}{D}S^{-} \end{array}\right).
\label{eq:solution}
\end{equation}
The bandwidth can be taken into account by a convolution of the solutions $|E^{(+/-)}_{2\omega}(\omega)|^2$ with the fundamental light spectra. 

\section{Optical parameters employed for the simulation}
\label{appendixc}
The absorption of NiO in the visible and near-infrared range is strongly affected by impurities such as excess oxygen\cite{NiOoptprop}. In order to determine an absorption spectrum for the simulation that suits the NiO crystals measured in our experiments, we used a polynomial fit to a known spectrum\cite{NiOoptprop} $\alpha_0(\omega)$ and adjusted for the specific impurity concentration by fitting $\alpha(\omega) = a \alpha_0(\omega) + b$ to a set of three direct absorption measurements that we carried out for $\lambda = 600$, $800$ and $1200$ nm.
These measurements also yielded a real part of the refractive index of about $n = 2.35\pm 0.05$. Due to the limited surface quality of our samples, we were not able to measure the dispersion in a static experiment with sufficient precision to predict the observed frequencies in the SHG response. \\

\begin{table}[htbp]
	\centering
		\begin{tabular}{cccccc}
		\hline
		\hline
			 $B_1$ \hspace{2mm} & $B_2$ \hspace{2mm} & $C_1$ $[$\textmu m$^2]$\hspace{2mm}  & $C_2$ $[$\textmu m$^2]$\hspace{2mm}  & $(\frac{\partial \rm{n}}{\partial \eta_{33}})$/$\rm{n}_0$ \hspace{2mm} & $(\frac{\partial \chi^{(2)}}{\partial \eta_{33}}) $/$ \chi^{(2)}_{0}$ \\
			\hline
			$1.22$ \vspace{2mm}  & $ 0.1 $ & $0.01$ & $0.152$ & $-1$ & $-10^3$ \\ 
			\hline
			\hline
		\end{tabular}
	\caption{Optical parameters used in the simulation. The Sellmeier coefficients are given in $B_1$ to $C_2$. The strain dependencies are shown normalized to the corresponding equilibrium quantity. 
The value of $(\frac{\partial \rm{n}}{\partial \eta_{33}})$/$\rm{n}_0$ matches experimental observations \cite{StrainNiO} without taking into account the observed wavelength dependence \cite{BoscoNiO}. 
The magnitude of $(\frac{\partial \chi^{(2)}}{\partial \eta_{33}})/\chi^{(2)}_{0}$ was chosen to best fit the data and corresponds to local changes in $\chi^{(2)}(z)$ of several percent. }
	\label{tab:optparameters}
\end{table}

\bibliography{apsnio}

\begin{thebibliography}{29}%
\makeatletter
\providecommand \@ifxundefined [1]{%
 \@ifx{#1\undefined}
}%
\providecommand \@ifnum [1]{%
 \ifnum #1\expandafter \@firstoftwo
 \else \expandafter \@secondoftwo
 \fi
}%
\providecommand \@ifx [1]{%
 \ifx #1\expandafter \@firstoftwo
 \else \expandafter \@secondoftwo
 \fi
}%
\providecommand \natexlab [1]{#1}%
\providecommand \enquote  [1]{``#1''}%
\providecommand \bibnamefont  [1]{#1}%
\providecommand \bibfnamefont [1]{#1}%
\providecommand \citenamefont [1]{#1}%
\providecommand \href@noop [0]{\@secondoftwo}%
\providecommand \href [0]{\begingroup \@sanitize@url \@href}%
\providecommand \@href[1]{\@@startlink{#1}\@@href}%
\providecommand \@@href[1]{\endgroup#1\@@endlink}%
\providecommand \@sanitize@url [0]{\catcode `\\12\catcode `\$12\catcode
  `\&12\catcode `\#12\catcode `\^12\catcode `\_12\catcode `\%12\relax}%
\providecommand \@@startlink[1]{}%
\providecommand \@@endlink[0]{}%
\providecommand \url  [0]{\begingroup\@sanitize@url \@url }%
\providecommand \@url [1]{\endgroup\@href {#1}{\urlprefix }}%
\providecommand \urlprefix  [0]{URL }%
\providecommand \Eprint [0]{\href }%
\providecommand \doibase [0]{http://dx.doi.org/}%
\providecommand \selectlanguage [0]{\@gobble}%
\providecommand \bibinfo  [0]{\@secondoftwo}%
\providecommand \bibfield  [0]{\@secondoftwo}%
\providecommand \translation [1]{[#1]}%
\providecommand \BibitemOpen [0]{}%
\providecommand \bibitemStop [0]{}%
\providecommand \bibitemNoStop [0]{.\EOS\space}%
\providecommand \EOS [0]{\spacefactor3000\relax}%
\providecommand \BibitemShut  [1]{\csname bibitem#1\endcsname}%
\let\auto@bib@innerbib\@empty
\bibitem [{\citenamefont {Kittel}(1948)}]{fmresonance}%
  \BibitemOpen
  \bibfield  {author} {\bibinfo {author} {\bibfnamefont {C.}~\bibnamefont
  {Kittel}},\ }\href {\doibase 10.1103/PhysRev.73.155} {\bibfield  {journal}
  {\bibinfo  {journal} {Phys. Rev.}\ }\textbf {\bibinfo {volume} {73}},\
  \bibinfo {pages} {155} (\bibinfo {year} {1948})}\BibitemShut {NoStop}%
\bibitem [{\citenamefont {Kittel}(1951)}]{afmresonance}%
  \BibitemOpen
  \bibfield  {author} {\bibinfo {author} {\bibfnamefont {C.}~\bibnamefont
  {Kittel}},\ }\href {\doibase 10.1103/PhysRev.82.565} {\bibfield  {journal}
  {\bibinfo  {journal} {Phys. Rev.}\ }\textbf {\bibinfo {volume} {82}},\
  \bibinfo {pages} {565} (\bibinfo {year} {1951})}\BibitemShut {NoStop}%
\bibitem [{\citenamefont {Kimel}\ \emph {et~al.}(2004)\citenamefont {Kimel},
  \citenamefont {Kirilyuk}, \citenamefont {Tsvetkov}, \citenamefont {Pisarev},\
  and\ \citenamefont {Rasing}}]{RasingTmFeO3}%
  \BibitemOpen
  \bibfield  {author} {\bibinfo {author} {\bibfnamefont {A.}~\bibnamefont
  {Kimel}}, \bibinfo {author} {\bibfnamefont {A.}~\bibnamefont {Kirilyuk}},
  \bibinfo {author} {\bibfnamefont {A.}~\bibnamefont {Tsvetkov}}, \bibinfo
  {author} {\bibfnamefont {R.}~\bibnamefont {Pisarev}}, \ and\ \bibinfo
  {author} {\bibfnamefont {T.}~\bibnamefont {Rasing}},\ }\href
  {http://www.scopus.com/inward/record.url?eid=2-s2.0-3042675283&partnerID=40&md5=359af6680a048f4d5a82d9d657b1784d}
  {\bibfield  {journal} {\bibinfo  {journal} {Nature}\ }\textbf {\bibinfo
  {volume} {429}},\ \bibinfo {pages} {850} (\bibinfo {year}
  {2004})}\BibitemShut {NoStop}%
\bibitem [{\citenamefont {Meiklejohn}\ and\ \citenamefont
  {Bean}(1957)}]{exchangebias}%
  \BibitemOpen
  \bibfield  {author} {\bibinfo {author} {\bibfnamefont {W.~H.}\ \bibnamefont
  {Meiklejohn}}\ and\ \bibinfo {author} {\bibfnamefont {C.~P.}\ \bibnamefont
  {Bean}},\ }\href {\doibase 10.1103/PhysRev.105.904} {\bibfield  {journal}
  {\bibinfo  {journal} {Phys. Rev.}\ }\textbf {\bibinfo {volume} {105}},\
  \bibinfo {pages} {904} (\bibinfo {year} {1957})}\BibitemShut {NoStop}%
\bibitem [{\citenamefont {Duong}\ \emph {et~al.}(2004)\citenamefont {Duong},
  \citenamefont {Satoh},\ and\ \citenamefont {Fiebig}}]{SHGpumpprobe}%
  \BibitemOpen
  \bibfield  {author} {\bibinfo {author} {\bibfnamefont {N.~P.}\ \bibnamefont
  {Duong}}, \bibinfo {author} {\bibfnamefont {T.}~\bibnamefont {Satoh}}, \ and\
  \bibinfo {author} {\bibfnamefont {M.}~\bibnamefont {Fiebig}},\ }\href
  {\doibase 10.1103/PhysRevLett.93.117402} {\bibfield  {journal} {\bibinfo
  {journal} {Phys. Rev. Lett.}\ }\textbf {\bibinfo {volume} {93}},\ \bibinfo
  {pages} {117402} (\bibinfo {year} {2004})}\BibitemShut {NoStop}%
\bibitem [{\citenamefont {Satoh}\ \emph {et~al.}(2006)\citenamefont {Satoh},
  \citenamefont {Duong},\ and\ \citenamefont {Fiebig}}]{SHGpumpprobe2}%
  \BibitemOpen
  \bibfield  {author} {\bibinfo {author} {\bibfnamefont {T.}~\bibnamefont
  {Satoh}}, \bibinfo {author} {\bibfnamefont {N.~P.}\ \bibnamefont {Duong}}, \
  and\ \bibinfo {author} {\bibfnamefont {M.}~\bibnamefont {Fiebig}},\ }\href
  {\doibase 10.1103/PhysRevB.74.012404} {\bibfield  {journal} {\bibinfo
  {journal} {Phys. Rev. B}\ }\textbf {\bibinfo {volume} {74}},\ \bibinfo
  {pages} {012404} (\bibinfo {year} {2006})}\BibitemShut {NoStop}%
\bibitem [{\citenamefont {Rubano}\ \emph {et~al.}(2010)\citenamefont {Rubano},
  \citenamefont {Satoh}, \citenamefont {Kimel}, \citenamefont {Kirilyuk},
  \citenamefont {Rasing},\ and\ \citenamefont {Fiebig}}]{SHGpulselength}%
  \BibitemOpen
  \bibfield  {author} {\bibinfo {author} {\bibfnamefont {A.}~\bibnamefont
  {Rubano}}, \bibinfo {author} {\bibfnamefont {T.}~\bibnamefont {Satoh}},
  \bibinfo {author} {\bibfnamefont {A.}~\bibnamefont {Kimel}}, \bibinfo
  {author} {\bibfnamefont {A.}~\bibnamefont {Kirilyuk}}, \bibinfo {author}
  {\bibfnamefont {T.}~\bibnamefont {Rasing}}, \ and\ \bibinfo {author}
  {\bibfnamefont {M.}~\bibnamefont {Fiebig}},\ }\href {\doibase
  10.1103/PhysRevB.82.174431} {\bibfield  {journal} {\bibinfo  {journal} {Phys.
  Rev. B}\ }\textbf {\bibinfo {volume} {82}},\ \bibinfo {pages} {174431}
  (\bibinfo {year} {2010})}\BibitemShut {NoStop}%
\bibitem [{\citenamefont {Hutchings}\ and\ \citenamefont
  {Samuelsen}(1972)}]{neutrondiff}%
  \BibitemOpen
  \bibfield  {author} {\bibinfo {author} {\bibfnamefont {M.~T.}\ \bibnamefont
  {Hutchings}}\ and\ \bibinfo {author} {\bibfnamefont {E.~J.}\ \bibnamefont
  {Samuelsen}},\ }\href {\doibase 10.1103/PhysRevB.6.3447} {\bibfield
  {journal} {\bibinfo  {journal} {Phys. Rev. B}\ }\textbf {\bibinfo {volume}
  {6}},\ \bibinfo {pages} {3447} (\bibinfo {year} {1972})}\BibitemShut
  {NoStop}%
\bibitem [{\citenamefont {Fiebig}\ \emph {et~al.}(2001)\citenamefont {Fiebig},
  \citenamefont {Fr\"ohlich}, \citenamefont {Lottermoser}, \citenamefont
  {Pavlov}, \citenamefont {Pisarev},\ and\ \citenamefont {Weber}}]{SHGinNiO}%
  \BibitemOpen
  \bibfield  {author} {\bibinfo {author} {\bibfnamefont {M.}~\bibnamefont
  {Fiebig}}, \bibinfo {author} {\bibfnamefont {D.}~\bibnamefont {Fr\"ohlich}},
  \bibinfo {author} {\bibfnamefont {T.}~\bibnamefont {Lottermoser}}, \bibinfo
  {author} {\bibfnamefont {V.~V.}\ \bibnamefont {Pavlov}}, \bibinfo {author}
  {\bibfnamefont {R.~V.}\ \bibnamefont {Pisarev}}, \ and\ \bibinfo {author}
  {\bibfnamefont {H.-J.}\ \bibnamefont {Weber}},\ }\href {\doibase
  10.1103/PhysRevLett.87.137202} {\bibfield  {journal} {\bibinfo  {journal}
  {Phys. Rev. Lett.}\ }\textbf {\bibinfo {volume} {87}},\ \bibinfo {pages}
  {137202} (\bibinfo {year} {2001})}\BibitemShut {NoStop}%
\bibitem [{\citenamefont {Thomsen}\ \emph {et~al.}(1986)\citenamefont
  {Thomsen}, \citenamefont {Grahn}, \citenamefont {Maris},\ and\ \citenamefont
  {Tauc}}]{StrainWave}%
  \BibitemOpen
  \bibfield  {author} {\bibinfo {author} {\bibfnamefont {C.}~\bibnamefont
  {Thomsen}}, \bibinfo {author} {\bibfnamefont {H.~T.}\ \bibnamefont {Grahn}},
  \bibinfo {author} {\bibfnamefont {H.~J.}\ \bibnamefont {Maris}}, \ and\
  \bibinfo {author} {\bibfnamefont {J.}~\bibnamefont {Tauc}},\ }\href {\doibase
  10.1103/PhysRevB.34.4129} {\bibfield  {journal} {\bibinfo  {journal} {Phys.
  Rev. B}\ }\textbf {\bibinfo {volume} {34}},\ \bibinfo {pages} {4129}
  (\bibinfo {year} {1986})}\BibitemShut {NoStop}%
\bibitem [{\citenamefont {Bosco}\ \emph {et~al.}(2002)\citenamefont {Bosco},
  \citenamefont {Azevedo},\ and\ \citenamefont {Acioli}}]{BoscoNiO}%
  \BibitemOpen
  \bibfield  {author} {\bibinfo {author} {\bibfnamefont {C.~A.~C.}\
  \bibnamefont {Bosco}}, \bibinfo {author} {\bibfnamefont {A.}~\bibnamefont
  {Azevedo}}, \ and\ \bibinfo {author} {\bibfnamefont {L.~H.}\ \bibnamefont
  {Acioli}},\ }\href {\doibase 10.1103/PhysRevB.66.125406} {\bibfield
  {journal} {\bibinfo  {journal} {Phys. Rev. B}\ }\textbf {\bibinfo {volume}
  {66}},\ \bibinfo {pages} {125406} (\bibinfo {year} {2002})}\BibitemShut
  {NoStop}%
\bibitem [{\citenamefont {Takahara}\ \emph {et~al.}(2012)\citenamefont
  {Takahara}, \citenamefont {Jinn}, \citenamefont {Wakabayashi}, \citenamefont
  {Moriyasu},\ and\ \citenamefont {Kohmoto}}]{StrainNiO}%
  \BibitemOpen
  \bibfield  {author} {\bibinfo {author} {\bibfnamefont {M.}~\bibnamefont
  {Takahara}}, \bibinfo {author} {\bibfnamefont {H.}~\bibnamefont {Jinn}},
  \bibinfo {author} {\bibfnamefont {S.}~\bibnamefont {Wakabayashi}}, \bibinfo
  {author} {\bibfnamefont {T.}~\bibnamefont {Moriyasu}}, \ and\ \bibinfo
  {author} {\bibfnamefont {T.}~\bibnamefont {Kohmoto}},\ }\href {\doibase
  10.1103/PhysRevB.86.094301} {\bibfield  {journal} {\bibinfo  {journal} {Phys.
  Rev. B}\ }\textbf {\bibinfo {volume} {86}},\ \bibinfo {pages} {094301}
  (\bibinfo {year} {2012})}\BibitemShut {NoStop}%
\bibitem [{\citenamefont {Roth}(1960)}]{roth}%
  \BibitemOpen
  \bibfield  {author} {\bibinfo {author} {\bibfnamefont {W.~L.}\ \bibnamefont
  {Roth}},\ }\href {\doibase http://dx.doi.org/10.1063/1.1735486} {\bibfield
  {journal} {\bibinfo  {journal} {Journal of Applied Physics}\ }\textbf
  {\bibinfo {volume} {31}},\ \bibinfo {pages} {2000} (\bibinfo {year}
  {1960})}\BibitemShut {NoStop}%
\bibitem [{\citenamefont {S\"anger}\ \emph {et~al.}(2006)\citenamefont
  {S\"anger}, \citenamefont {Pavlov}, \citenamefont {Bayer},\ and\
  \citenamefont {Fiebig}}]{SHGdomains}%
  \BibitemOpen
  \bibfield  {author} {\bibinfo {author} {\bibfnamefont {I.}~\bibnamefont
  {S\"anger}}, \bibinfo {author} {\bibfnamefont {V.~V.}\ \bibnamefont
  {Pavlov}}, \bibinfo {author} {\bibfnamefont {M.}~\bibnamefont {Bayer}}, \
  and\ \bibinfo {author} {\bibfnamefont {M.}~\bibnamefont {Fiebig}},\ }\href
  {\doibase 10.1103/PhysRevB.74.144401} {\bibfield  {journal} {\bibinfo
  {journal} {Phys. Rev. B}\ }\textbf {\bibinfo {volume} {74}},\ \bibinfo
  {pages} {144401} (\bibinfo {year} {2006})}\BibitemShut {NoStop}%
\bibitem [{\citenamefont {Newman}\ and\ \citenamefont
  {Chrenko}(1959)}]{NiOoptprop}%
  \BibitemOpen
  \bibfield  {author} {\bibinfo {author} {\bibfnamefont {R.}~\bibnamefont
  {Newman}}\ and\ \bibinfo {author} {\bibfnamefont {R.~M.}\ \bibnamefont
  {Chrenko}},\ }\href {\doibase 10.1103/PhysRev.114.1507} {\bibfield  {journal}
  {\bibinfo  {journal} {Phys. Rev.}\ }\textbf {\bibinfo {volume} {114}},\
  \bibinfo {pages} {1507} (\bibinfo {year} {1959})}\BibitemShut {NoStop}%
\bibitem [{\citenamefont {Fernandez}\ \emph {et~al.}(1998)\citenamefont
  {Fernandez}, \citenamefont {Vettier}, \citenamefont {de~Bergevin},
  \citenamefont {Giles},\ and\ \citenamefont {Neubeck}}]{bergevinNiO}%
  \BibitemOpen
  \bibfield  {author} {\bibinfo {author} {\bibfnamefont {V.}~\bibnamefont
  {Fernandez}}, \bibinfo {author} {\bibfnamefont {C.}~\bibnamefont {Vettier}},
  \bibinfo {author} {\bibfnamefont {F.}~\bibnamefont {de~Bergevin}}, \bibinfo
  {author} {\bibfnamefont {C.}~\bibnamefont {Giles}}, \ and\ \bibinfo {author}
  {\bibfnamefont {W.}~\bibnamefont {Neubeck}},\ }\href {\doibase
  10.1103/PhysRevB.57.7870} {\bibfield  {journal} {\bibinfo  {journal} {Phys.
  Rev. B}\ }\textbf {\bibinfo {volume} {57}},\ \bibinfo {pages} {7870}
  (\bibinfo {year} {1998})}\BibitemShut {NoStop}%
\bibitem [{\citenamefont {Blume}\ and\ \citenamefont
  {Gibbs}(1988)}]{blumegibbs}%
  \BibitemOpen
  \bibfield  {author} {\bibinfo {author} {\bibfnamefont {M.}~\bibnamefont
  {Blume}}\ and\ \bibinfo {author} {\bibfnamefont {D.}~\bibnamefont {Gibbs}},\
  }\href {\doibase 10.1103/PhysRevB.37.1779} {\bibfield  {journal} {\bibinfo
  {journal} {Phys. Rev. B}\ }\textbf {\bibinfo {volume} {37}},\ \bibinfo
  {pages} {1779} (\bibinfo {year} {1988})}\BibitemShut {NoStop}%
\bibitem [{\citenamefont {Bergevin}\ and\ \citenamefont
  {Brunel}(1972)}]{bergevin1972}%
  \BibitemOpen
  \bibfield  {author} {\bibinfo {author} {\bibfnamefont {F.~D.}\ \bibnamefont
  {Bergevin}}\ and\ \bibinfo {author} {\bibfnamefont {M.}~\bibnamefont
  {Brunel}},\ }\href {\doibase http://dx.doi.org/10.1016/0375-9601(72)91054-7}
  {\bibfield  {journal} {\bibinfo  {journal} {Physics Letters A}\ }\textbf
  {\bibinfo {volume} {39}},\ \bibinfo {pages} {141 } (\bibinfo {year}
  {1972})}\BibitemShut {NoStop}%
\bibitem [{\citenamefont {Tanaka}\ and\ \citenamefont {et~al.}(2012)}]{sacla}%
  \BibitemOpen
  \bibfield  {author} {\bibinfo {author} {\bibfnamefont {H.}~\bibnamefont
  {Tanaka}}\ and\ \bibinfo {author} {\bibfnamefont {M.~Y.}\ \bibnamefont
  {et~al.}},\ }\href {\doibase 10.1038/nphoton.2012.141} {\bibfield  {journal}
  {\bibinfo  {journal} {Nature Photonics}\ }\textbf {\bibinfo {volume} {6}},\
  \bibinfo {pages} {540–544} (\bibinfo {year} {2012})}\BibitemShut {NoStop}%
\bibitem [{\citenamefont {Lehmk\"uhler}\ \emph {et~al.}(2014)\citenamefont
  {Lehmk\"uhler}, \citenamefont {Gutt}, \citenamefont {Fischer}, \citenamefont
  {Schroer}, \citenamefont {Sikorski}, \citenamefont {Song}, \citenamefont
  {Roseker}, \citenamefont {Glownia}, \citenamefont {Chollet}, \citenamefont
  {Nelson}, \citenamefont {Tono}, \citenamefont {Katayama}, \citenamefont
  {Yabashi}, \citenamefont {Ishikawa}, \citenamefont {Robert},\ and\
  \citenamefont {Gr\"ubel}}]{saclapulselength}%
  \BibitemOpen
  \bibfield  {author} {\bibinfo {author} {\bibfnamefont {F.}~\bibnamefont
  {Lehmk\"uhler}}, \bibinfo {author} {\bibfnamefont {C.}~\bibnamefont {Gutt}},
  \bibinfo {author} {\bibfnamefont {B.}~\bibnamefont {Fischer}}, \bibinfo
  {author} {\bibfnamefont {M.~A.}\ \bibnamefont {Schroer}}, \bibinfo {author}
  {\bibfnamefont {M.}~\bibnamefont {Sikorski}}, \bibinfo {author}
  {\bibfnamefont {S.}~\bibnamefont {Song}}, \bibinfo {author} {\bibfnamefont
  {W.}~\bibnamefont {Roseker}}, \bibinfo {author} {\bibfnamefont
  {J.}~\bibnamefont {Glownia}}, \bibinfo {author} {\bibfnamefont
  {M.}~\bibnamefont {Chollet}}, \bibinfo {author} {\bibfnamefont
  {S.}~\bibnamefont {Nelson}}, \bibinfo {author} {\bibfnamefont
  {K.}~\bibnamefont {Tono}}, \bibinfo {author} {\bibfnamefont {T.}~\bibnamefont
  {Katayama}}, \bibinfo {author} {\bibfnamefont {M.}~\bibnamefont {Yabashi}},
  \bibinfo {author} {\bibfnamefont {T.}~\bibnamefont {Ishikawa}}, \bibinfo
  {author} {\bibfnamefont {A.}~\bibnamefont {Robert}}, \ and\ \bibinfo {author}
  {\bibfnamefont {G.}~\bibnamefont {Gr\"ubel}},\ }\href {\doibase
  10.1038/srep05234} {\bibfield  {journal} {\bibinfo  {journal} {Scientific
  Reports}\ }\textbf {\bibinfo {volume} {4}} (\bibinfo {year} {2014}),\
  10.1038/srep05234}\BibitemShut {NoStop}%
\bibitem [{\citenamefont {Jurani\'{c}}\ \emph {et~al.}(2014)\citenamefont
  {Jurani\'{c}}, \citenamefont {Stepanov}, \citenamefont {Ischebeck},
  \citenamefont {Schlott}, \citenamefont {Pradervand}, \citenamefont {Patthey},
  \citenamefont {Radovi\'{c}}, \citenamefont {Gorgisyan}, \citenamefont
  {Rivkin}, \citenamefont {Hauri}, \citenamefont {Monoszlai}, \citenamefont
  {Ivanov}, \citenamefont {Peier}, \citenamefont {Liu}, \citenamefont
  {Togashi}, \citenamefont {Owada}, \citenamefont {Ogawa}, \citenamefont
  {Katayama}, \citenamefont {Yabashi},\ and\ \citenamefont
  {Abela}}]{saclatiming}%
  \BibitemOpen
  \bibfield  {author} {\bibinfo {author} {\bibfnamefont {P.~N.}\ \bibnamefont
  {Jurani\'{c}}}, \bibinfo {author} {\bibfnamefont {A.}~\bibnamefont
  {Stepanov}}, \bibinfo {author} {\bibfnamefont {R.}~\bibnamefont {Ischebeck}},
  \bibinfo {author} {\bibfnamefont {V.}~\bibnamefont {Schlott}}, \bibinfo
  {author} {\bibfnamefont {C.}~\bibnamefont {Pradervand}}, \bibinfo {author}
  {\bibfnamefont {L.}~\bibnamefont {Patthey}}, \bibinfo {author} {\bibfnamefont
  {M.}~\bibnamefont {Radovi\'{c}}}, \bibinfo {author} {\bibfnamefont
  {I.}~\bibnamefont {Gorgisyan}}, \bibinfo {author} {\bibfnamefont
  {L.}~\bibnamefont {Rivkin}}, \bibinfo {author} {\bibfnamefont {C.~P.}\
  \bibnamefont {Hauri}}, \bibinfo {author} {\bibfnamefont {B.}~\bibnamefont
  {Monoszlai}}, \bibinfo {author} {\bibfnamefont {R.}~\bibnamefont {Ivanov}},
  \bibinfo {author} {\bibfnamefont {P.}~\bibnamefont {Peier}}, \bibinfo
  {author} {\bibfnamefont {J.}~\bibnamefont {Liu}}, \bibinfo {author}
  {\bibfnamefont {T.}~\bibnamefont {Togashi}}, \bibinfo {author} {\bibfnamefont
  {S.}~\bibnamefont {Owada}}, \bibinfo {author} {\bibfnamefont
  {K.}~\bibnamefont {Ogawa}}, \bibinfo {author} {\bibfnamefont
  {T.}~\bibnamefont {Katayama}}, \bibinfo {author} {\bibfnamefont
  {M.}~\bibnamefont {Yabashi}}, \ and\ \bibinfo {author} {\bibfnamefont
  {R.}~\bibnamefont {Abela}},\ }\href {\doibase 10.1364/OE.22.030004}
  {\bibfield  {journal} {\bibinfo  {journal} {Opt. Express}\ }\textbf {\bibinfo
  {volume} {22}},\ \bibinfo {pages} {30004} (\bibinfo {year}
  {2014})}\BibitemShut {NoStop}%
\bibitem [{\citenamefont {Durbin}\ \emph {et~al.}(2012)\citenamefont {Durbin},
  \citenamefont {Clevenger}, \citenamefont {Graber},\ and\ \citenamefont
  {Henning}}]{GaAs}%
  \BibitemOpen
  \bibfield  {author} {\bibinfo {author} {\bibfnamefont {S.~M.}\ \bibnamefont
  {Durbin}}, \bibinfo {author} {\bibfnamefont {T.}~\bibnamefont {Clevenger}},
  \bibinfo {author} {\bibfnamefont {T.}~\bibnamefont {Graber}}, \ and\ \bibinfo
  {author} {\bibfnamefont {R.}~\bibnamefont {Henning}},\ }\href {\doibase
  10.1038/NPHOTON.2011.327} {\bibfield  {journal} {\bibinfo  {journal} {Nature
  Photonics}\ }\textbf {\bibinfo {volume} {6}},\ \bibinfo {pages} {111}
  (\bibinfo {year} {2012})}\BibitemShut {NoStop}%
\bibitem [{\citenamefont {Jifang}\ \emph {et~al.}(1991)\citenamefont {Jifang},
  \citenamefont {Fisher},\ and\ \citenamefont {Manghnzmi}}]{elastic}%
  \BibitemOpen
  \bibfield  {author} {\bibinfo {author} {\bibfnamefont {W.}~\bibnamefont
  {Jifang}}, \bibinfo {author} {\bibfnamefont {E.~S.}\ \bibnamefont {Fisher}},
  \ and\ \bibinfo {author} {\bibfnamefont {M.~H.}\ \bibnamefont {Manghnzmi}},\
  }\href {http://stacks.iop.org/0256-307X/8/i=3/a=012} {\bibfield  {journal}
  {\bibinfo  {journal} {Chinese Physics Letters}\ }\textbf {\bibinfo {volume}
  {8}},\ \bibinfo {pages} {153} (\bibinfo {year} {1991})}\BibitemShut {NoStop}%
\bibitem [{\citenamefont {Satoh}\ \emph {et~al.}(2007)\citenamefont {Satoh},
  \citenamefont {Aken}, \citenamefont {Duong}, \citenamefont {Lottermoser},\
  and\ \citenamefont {Fiebig}}]{demagCr2O3}%
  \BibitemOpen
  \bibfield  {author} {\bibinfo {author} {\bibfnamefont {T.}~\bibnamefont
  {Satoh}}, \bibinfo {author} {\bibfnamefont {B.~B.~V.}\ \bibnamefont {Aken}},
  \bibinfo {author} {\bibfnamefont {N.~P.}\ \bibnamefont {Duong}}, \bibinfo
  {author} {\bibfnamefont {T.}~\bibnamefont {Lottermoser}}, \ and\ \bibinfo
  {author} {\bibfnamefont {M.}~\bibnamefont {Fiebig}},\ }\href {\doibase
  http://dx.doi.org/10.1016/j.jmmm.2006.10.630} {\bibfield  {journal} {\bibinfo
   {journal} {Journal of Magnetism and Magnetic Materials}\ }\textbf {\bibinfo
  {volume} {310}},\ \bibinfo {pages} {1604 } (\bibinfo {year} {2007})},\
  \bibinfo {note} {proceedings of the 17th International Conference on
  Magnetism The International Conference on Magnetism}\BibitemShut {NoStop}%
\bibitem [{\citenamefont {Chatterji}\ \emph {et~al.}(2009)\citenamefont
  {Chatterji}, \citenamefont {McIntyre},\ and\ \citenamefont
  {Lindgard}}]{criticalexp}%
  \BibitemOpen
  \bibfield  {author} {\bibinfo {author} {\bibfnamefont {T.}~\bibnamefont
  {Chatterji}}, \bibinfo {author} {\bibfnamefont {G.~J.}\ \bibnamefont
  {McIntyre}}, \ and\ \bibinfo {author} {\bibfnamefont {P.-A.}\ \bibnamefont
  {Lindgard}},\ }\href {\doibase 10.1103/PhysRevB.79.172403} {\bibfield
  {journal} {\bibinfo  {journal} {Phys. Rev. B}\ }\textbf {\bibinfo {volume}
  {79}},\ \bibinfo {pages} {172403} (\bibinfo {year} {2009})}\BibitemShut
  {NoStop}%
\bibitem [{\citenamefont {Mandal}\ \emph {et~al.}(2009)\citenamefont {Mandal},
  \citenamefont {Menon}, \citenamefont {Maccherozzi},\ and\ \citenamefont
  {Belkhou}}]{NiOstrainreorient}%
  \BibitemOpen
  \bibfield  {author} {\bibinfo {author} {\bibfnamefont {S.}~\bibnamefont
  {Mandal}}, \bibinfo {author} {\bibfnamefont {K.~S.~R.}\ \bibnamefont
  {Menon}}, \bibinfo {author} {\bibfnamefont {F.}~\bibnamefont {Maccherozzi}},
  \ and\ \bibinfo {author} {\bibfnamefont {R.}~\bibnamefont {Belkhou}},\ }\href
  {\doibase 10.1103/PhysRevB.80.184408} {\bibfield  {journal} {\bibinfo
  {journal} {Phys. Rev. B}\ }\textbf {\bibinfo {volume} {80}},\ \bibinfo
  {pages} {184408} (\bibinfo {year} {2009})}\BibitemShut {NoStop}%
\bibitem [{\citenamefont {Clendenen}\ and\ \citenamefont
  {Drickamer}(1966)}]{expansion}%
  \BibitemOpen
  \bibfield  {author} {\bibinfo {author} {\bibfnamefont {R.~L.}\ \bibnamefont
  {Clendenen}}\ and\ \bibinfo {author} {\bibfnamefont {H.~G.}\ \bibnamefont
  {Drickamer}},\ }\href {\doibase http://dx.doi.org/10.1063/1.1726610}
  {\bibfield  {journal} {\bibinfo  {journal} {The Journal of Chemical Physics}\
  }\textbf {\bibinfo {volume} {44}},\ \bibinfo {pages} {4223} (\bibinfo {year}
  {1966})}\BibitemShut {NoStop}%
\bibitem [{\citenamefont {Seltz}\ \emph {et~al.}(1940)\citenamefont {Seltz},
  \citenamefont {DeWitt},\ and\ \citenamefont {McDonald}}]{heatcapacity}%
  \BibitemOpen
  \bibfield  {author} {\bibinfo {author} {\bibfnamefont {H.}~\bibnamefont
  {Seltz}}, \bibinfo {author} {\bibfnamefont {B.~J.}\ \bibnamefont {DeWitt}}, \
  and\ \bibinfo {author} {\bibfnamefont {H.~J.}\ \bibnamefont {McDonald}},\
  }\href {\doibase 10.1021/ja01858a022} {\bibfield  {journal} {\bibinfo
  {journal} {Journal of the American Chemical Society}\ }\textbf {\bibinfo
  {volume} {62}},\ \bibinfo {pages} {88} (\bibinfo {year} {1940})}\BibitemShut
  {NoStop}%
\bibitem [{\citenamefont {Born}\ and\ \citenamefont {Wolf}(1964)}]{bornwolf}%
  \BibitemOpen
  \bibfield  {author} {\bibinfo {author} {\bibfnamefont {M.}~\bibnamefont
  {Born}}\ and\ \bibinfo {author} {\bibfnamefont {E.}~\bibnamefont {Wolf}},\
  }\href@noop {} {\emph {\bibinfo {title} {Principles of optics:
  electromagnetic theory of propagation, interference and diffraction of
  light}}}\ (\bibinfo  {publisher} {Oxford, Pergamon Press},\ \bibinfo {year}
  {1964})\BibitemShut {NoStop}%
\end{thebibliography}%

\end{document}